\documentclass[linenumber]{aastex631}
\usepackage{CJK}

\shorttitle{$L-T-E$ correlation}
\shortauthors{Deng et al.}
\graphicspath{{./}{figures/}}
\usepackage{amsmath}
\usepackage{longtable}
\makeatletter

\newcommand{\Rmnum}[1]{\expandafter\@slowromancap\romannumeral #1@}
\makeatother
\begin{document}
\begin{CJK*}{UTF8}{gbsn}
\title{Pseudo Redshifts of Gamma-Ray Bursts Derived from
the $L-T-E$ Correlation}
\correspondingauthor{Yong-Feng Huang}
\email{hyf@nju.edu.cn}
\author{Chen Deng (邓晨)}
\affiliation{School of Astronomy and Space Science, Nanjing
University\\ Nanjing 210023, People's Republic of China}

\author{Yong-Feng Huang (黄永锋)}
\affiliation{School of Astronomy and Space Science, Nanjing
University\\ Nanjing 210023, People's Republic of China}
\affiliation{Key Laboratory of Modern Astronomy and Astrophysics
(Nanjing University)\\ Ministry of Education, People's Republic of
China}

\author{Fan Xu (许帆)}
\affiliation{School of Astronomy and Space Science, Nanjing
University\\ Nanjing 210023, People's Republic of China}

\begin{abstract}
The X-ray afterglow of many gamma-ray bursts (GRBs) exhibits a
plateau phase before the normal power-law decay stage, which may
be related to continued activities of the central engine.
\cite{Tang2019} collected 174 such GRBs and confirmed the so
called $L-T-E$ correlation which involves three key parameters,
i.e., the isotropic $\gamma$-ray energy $E_{\gamma,\rm iso}$ of
the prompt phase, the end time $T_{a}$ of the plateau phase and
the corresponding X-ray luminosity $L_{X}$. In this study, the
$L-T-E$ correlation is confirmed and updated as $L_{X} \propto
T_{a}^{-0.99} E_{\gamma ,\rm iso}^{0.86}$ with a large sample
consisting of 210 plateau GRBs with known redshifts. The tight
correlation is then applied to derive the pseudo redshift of other
130 plateau GRBs whose redshifts are not directly measured.
Statistical analysis is also carried out on this pseudo redshift
sample.



\end{abstract}


\section{Introduction}
\label{sec:intro}

Gamma-ray bursts (GRBs) are one of the most powerful stellar
explosions in our Universe with the redshifts measured up to $z
\sim 9$ (\citealt{Cucchiara2011}; \citealt{Salvaterra2012}). In
view of this, GRBs are a hopeful tool to probe high-redshift
universe \citep{amati2013}. The spectra of GRBs are generally
nonthermal and can be fitted well with the so-called Band function
\citep{band1993}. Traditionally, GRBs can be roughly classified
into long ($T_{90}>2\rm~s$) and short ($T_{90}<2\rm~s$) categories
by their duration $T_{90}$ (defined as the time interval during
which 5$\%$ --- 95$\%$ of the prompt $\gamma$-ray photons are
detected) \citep{Kouveliotou1993}. Long GRBs typically have a
duration of 20 --- 30 s, while short GRBs usually last for 0.2 ---
0.3 s. It is widely believed that long GRBs originate from the
death of massive stars (\citealt{woosley1993};
\citealt{macfadyen1999}; \citealt{galama1998};
\citealt{hjorth2003}; \citealt{campana2006}), and short GRBs
originate from the merger of binary compact
stars (\citealt{Paczynski1986}; \citealt{Eichler1989};
\citealt{nakar2007}). A recent breakthrough in the field is the
detection by the advanced Laser Interferometer Gravitational Wave
Observatory of the gravitational-wave event GW 170817 from a
binary neutron star merger, which is in association with the short
burst GRB 170817A \citep{Abbott2017}. Nevertheless,
\cite{norris2006} found a class of short bursts with extended
emission that share some of the properties of both long and short
GRBs. Long-duration GRB 211211A, associated with a kilonova,
exhibits three components in the prompt phase, also challenging
our understanding of GRBs
\citep{Rastinejad2022arXiv220410864R,gao2022ApJl}. Note
that the duration of GRBs is related to the energy band of the
detector \citep{Bromberg2013}. In other words, the duration
measured by different detectors could be different even for the
same burst. Instead of phenomenological classification based on
the duration, \cite{zhang2006} suggested physically classifying
GRBs into two different categories, compact star type (type
\Rmnum{1}) and massive star type (type \Rmnum{2}).

The emission process of GRBs can be divided into two
phenomenological phases, namely, the prompt phase and the
afterglow phase. The ``fireball'' model is the most popular
mechanism engaged to explain the origin of prompt GRB emission
(\citealt{Rees1992mnras}; \citealt{Piran1993MNRAS} ;
\citealt{Wijers1997MNRAS}). The central engine, as a result of
either the death of a supermassive star or the merge of binary
neutron stars, releases a sequence of relativistic shells with
different velocities. The collisions of various shells accelerate
electrons to produce complex and rapidly changing light curve in
the prompt GRB phase, while the interaction of the combined
materials with the ambient medium generates the multi-band
afterglow at late stages (\citealt{meszaros1997ApJ, Waxman1997ApJ,
sari1997ApJ, sari1998ApJ, Panaitescu1998ApJ,huang1999MNRAS,
huang2000ApJ, Yi2020ApJ}). Thanks to the rapid response and
precise positioning of the Swift satellite, various intricate
features have been discovered in GRB afterglows
(\citealt{Burrows2005, Nousek2006apj, OBrien2006ApJ}). A portion
of the observed afterglow light curves favor the scenario
involving continued late-time energy injection by the central
engine (\citealt{Willingale2007ApJ}). As the observational data
accumulated, a picture of the canonical afterglow lightcurves
consisting of five segments was established
(\citealt{zhang2006ApJ}). The plateau phase usually follows the
steep decay phase and is characterized by a shallow temporal decay
index ($\sim$ 0), which is a common phenomenon in the Swift GRBs
(\citealt{xu2012,Tang2019}). The physical mechanism behind the plateau
phase may be the continuous activity of the central engine
(\citealt{kumar2008Sci,Cannizzo2009ApJ, xu2012, yi2022ApJ}). A shallow
decay phase followed by very steep decay phase with a power-law
temporal index $< -3$ is called an internal plateau
(\citealt{Nousek2006apj, dainotti2011ApJ, Dainotti2017NewAR,
xu2021ApJ}, which may be due to the switching off of the energy
injection after the supramassive neutron star collapses into a black
hole (\citealt{chen2007ApJ,hou2018ApJ}).

GRBs can serve as a probe of the distant universe because of their
extreme luminosities(\citealt{Ghirlanda2006NJPh}; 
\citealt{Cardone2009MNRAS,cardone2010MNRAS,Dainotti2013bMNRAS};
\citealt{amati2013}; \citealt{Postnikov2014ApJ};
\citealt{Dainotti2017NewAR,xu2021ApJ,hjp2021MNRAS,Muccino2021ApJ,wangfy2022ApJ}; 
\citealt{cao2022MNRAS,Hujp2022A&A}). Since GRBs may be
produced by more than one kind of central engines, the intrinsic
energy release of GRBs spans in a wide range, i.e., from $\sim
10^{48}$ erg up to $\sim 10^{55}$ erg (\citealt{kumar2015PhR}). As
a result, GRBs cannot directly serve as a standard candle.
Luckily, several tight empirical correlations have been
established among various observational parameters of GRBs, which
may effectively help alleviate this problem. These correlations
usually involve two or more quantities, and can be classified as
prompt parameter correlations (\citealt{norris2000ApJ,
amati2002A&A, Ghirlanda2004ApJ, Ghirlanda2006NJPh,
yonetoku2004ApJ, Willingale2007ApJ};
\citealt{Dainotti2018PASP,dainotti2018AdAst}), afterglow parameter
correlations (\citealt{dainotti2008MNRAS, dainotti2010ApJ,
Ghisellini2009MNRAS}), and prompt-afterglow correlations
(\citealt{liang2005ApJ,xu2012, Bernardini2012MNRAS, Izzo2015A&A,
dainotti2016ApJ,Dainotti2022MNRAS, Si2018ApJ, Tang2019,xu2021ApJ}). 
These powerful empirical correlations can not only
deepen our understanding of the physical mechanism of GRBs but
also act as theoretical model discriminators
(\citealt{Dainotti2017NewAR}; \citealt{Dainotti2018PASP}).

Most of the detected GRBs do not have a measured redshift, which
prevents their application in cosmology to some extent. It is
interesting to note that some authors have attempted to use
empirical correlations to derive pseudo redshifts of GRBs
(\citealt{Atteia2003A&A, yonetoku2004ApJ, dainotti2011ApJ,
Tsutsui2013MNRAS, Tan2013ApJ}). For example,
\cite{Tsutsui2013MNRAS} calculated the pseudo redshifts of 71
bright short GRBs using a tight $L_{\rm p} - E_{\rm p}$
correlation. More recently, \citet{Zhanggq2018ApJ} used the
$L_{\rm p}-E_{\rm p}$ correlation to explore the luminosity
function and formation rate of short GRBs.
In addition, some authors investigated the evolution
of the luminosity function and formation rate in detail for
short and long GRBs (\citealt{Wanderman2015MNRAS,
Paul2018MNRAS,Dainotti2021ApJL}). Following
\citet{Zhanggq2018ApJ}, \citet{Zitouni2018Ap&SS} estimated the
pseudo redshifts of 1017 Fermi GRBs and investigated their
intrinsic duration distribution in the GRB rest frame.

For GRBs with a plateau phase in the X-ray afterglow light curve,
\citet{dainotti2008MNRAS} firstly discovered an anti-relation
between the plateau end time $T_{\rm a}$ and the corresponding
X-ray luminosity $L_{\rm X}$ using 33 long GRBs detected by the
Swift satellite. Consequently, they used this $L_{\rm X}-T_{\rm
a}$ correlation to estimate pseudo redshifts of GRBs and pointed
out that the results are significantly affected by the uncertainty
of the observed quantities as well as the intrinsic dispersion of
the empirical correlation \citep{dainotti2011ApJ}. Subsequently,
the $L_{\rm X}-T_{\rm a}$ relation was extended to a
three-parameter correlation among $L_{\rm X}-T_{\rm a}-L_{\rm p}$
by using 40 golden GRBs, where $L_{\rm p}$ is the peak luminosity
of GRBs \citep{dainotti2016ApJ}. Furthermore, the platinum
sample consisting of 47 GRBs, as an improvement of the golden
sample in the previous study, was used to derive a tighter $L_{\rm
X}-T_{\rm a}-L_{\rm p}$ correlation
\citep{Dainotti2017bApJ,Dainotti2020bApJ}. More elaborate
problems in the $L_{\rm X}-T_{\rm a}$ correlation and its
extensions, such as selection effects and redshift evolution, have
been adequately discussed \citep{Dainotti2013aApJ,Dainotti2015bMNRAS,
Dainotti2015aApJ,Dainotti2017bApJ,Dainotti2017aA&A,
Dainotti2020bApJ,Dainotti2022MNRAS}. On the other hand,
\citet{xu2012} introduced
the prompt isotropic $\gamma$-ray energy ($E_{\gamma,\rm iso}$) as
a third key parameter and obtained a tight three-parameter
correlation among $L_{\rm X}-T_{\rm a}-E_{\gamma,\rm iso}$ (it is
called the $L-T-E$ correlation hereafter). \citet{Tang2019}
further collected a large sample consisting of 174 GRBs and
confirmed the existence of the $L-T-E$ correlation, e.g., $L_{\rm
X}\propto T_{\rm a}^{-1.01}E_{\gamma,\rm iso}^{0.85}$. It is
interesting to note that the sample of \citet{Tang2019} includes 6
short GRBs and 11 internal plateau GRBs. They all satisfy the same
$L-T-E$ correlation as long GRBs.

Recently, the $L-T-E$ correlation was further confirmed in an
independent study by \cite{Ding2022ApSS} with a slightly expanded
GRB sample. \cite{Izzo2015A&A} derived the Combo-relation as
$L_{\rm X}\propto E_{\rm p}^{0.84\pm 0.08}(T_{\rm
a}/|1+\alpha|)^{-1}$ and argued that there is no redshift
evolution in it. Subsequently, some authors proved that the
Combo-relation can be used to probe the universe
\citep{Luongo2020A&A,Luongo2021MNRAS,Muccino2021ApJ}.
\cite{Srinivasaragavan2020ApJ} utilized the plateau GRBs
to study the closure relations between the temporal and spectral
indices in the afterglow phase and showed that the external shock
model is generally favored in most cases. More interestingly,
there are also two-parameter correlations in the optical afterglow
plateau (\citealt{Zaninoni2013A&A,Dainotti2020aApJL}). Based on
these studies, some three-parameter correlations in optical
afterglow phase were established recently
(\citealt{Si2018ApJ,Dainotti2022ApJS}). For
comprehensive reviews on the statistics of the plateau
 phase in GRB afterglows, please refer to
\citet{Dainotti2017NewAR}, \citet{dainotti2018AdAst},
\citet{zhao2019ApJ} and \citet{Wangff2020ApJ}.

In this study, we will use the $L-T-E$ correlation to estimate the
redshifts of GRBs. The structure of our paper is organized as
follow. Section \ref{sec:sample} describes the sample collection
process and some necessary data analysis. The $L-T-E$ correlation
is then revisited for plateau GRBs with our sample in Section
\ref{sec:LTE}. Pseudo redshifts are calculated for a large number
of GRBs based on the $L-T-E$ correlation in Section
\ref{sec:estimator}, and the distributions of several derived
parameters are also analyzed. Finally, the
results are briefly summarized and discussed in Section
\ref{sec:summary}.

\section{sample selection and data analysis}
\label{sec:sample}

The shallow decay (plateau) phase is a common phenomenon in the
afterglow of GRBs. In this study, we concentrate on the plateau
phase in the X-ray afterglow of GRBs and attempt to utilize the
$L_{\rm X}-T_{\rm a}-E_{\gamma,\rm iso}$ correlation as a redshift
indicator. This so called $L-T-E$ correlation was originally
proposed by \citet{xu2012}. \citet{Tang2019} collected a large
sample including 174 GRBs with well defined redshifts and
confirmed that the correlation does exist. Here we will first
upgrade the $L-T-E$ correlation by using an updated GRB sample
containing as many events as possible, which is then utilized to
calculate the pseudo redshift of other plateau GRBs without a
distance measurement. For this purpose, we have constructed two
GRB samples, one with the redshift measured (designated as the
Starting Sample), and the other without redshift measurement
(designated as the Target Sample). In both samples, each GRB is
characterized by an obvious plateau phase in the X-ray afterglow
light curve.

The criteria used by \citet{Tang2019} to select the sample are
fairly strict, which is very important to clearly define a plateau
phase. Following almost the same criteria as \citet{Tang2019}, we
have re-examined all the GRBs observed by Swift between 2005 March
and 2022 May \citep{Evans2007A&A,Evans2009mnras}. To be more
specific, our criteria are: (1) there should be a clear plateau
phase in the X-ray afterglow light curve with the temporal

power-law decay index being in the range of $-1$ to 1; (2) Abundant
data points are available to well define the plateau phase. (3)
The redshifts of the GRBs samples should be well-defined so that
we can calculate the isotropic emission energy and luminosity. To
expand the sample size as far as possible, we have included those
GRBs with X-ray flares superposed on the plateau phase. In these
cases, the X-ray flares are simply omitted when fitting the light
curve of the plateau phase. As a result, our Starting Sample with
known redshift includes 210 GRBs. Comparing with the sample of
\citet{Tang2019}, 36 GRBs are new in our data set. For GRBs
without a redshift measurement, our Target Sample contains 130
events. Pseudo redshift will be calculated for these Target Sample
events based on the updated $L-T-E$ correlation. The X-ray light
curves of all the GRBs in our two samples have been downloaded
from the Swift light curve repository
\citep{Evans2007A&A,Evans2009mnras}.

The plateau phase and the subsequent normal decay phase are then
fitted by a smoothly broken power-law function, which is
\citep{yi2016ApJS,Tang2019}
\begin{eqnarray} \label{func1}
F_{X}(t) & = & F_{X 0}\left[\left(\frac{t}{T_{\rm a,obs}}\right)^{\alpha_{1}
\omega}+\left(\frac{t}{T_{\rm a,obs}}\right)^{\alpha_{2} \omega}\right]^{-1 /
\omega} ,
\end{eqnarray}
where $\alpha_{1}$, $\alpha_{2}$, $T_{\rm a,obs}$ and $\omega$
represent the temporal power-law decay index of the plateau phase,
the decay index of the post-break segment, the break time in the
observer's frame, and the smoothness factor of the transition
segment, respectively. The end time of the plateau in the GRB rest
frame can be obtained by $T_{\rm a}=T_{\rm a,obs}/(1+z)$. The flux
at the break time is $F_{X 0}\times 2^{-1/ \omega}$. For
simplicity, the smoothness factor is fixed as 1.0 in our fitting
process. Correspondingly, the flux at the end of the plateau phase
is $F_{X 0}/2$.

\begin{figure}[ht!]
\gridline{\fig{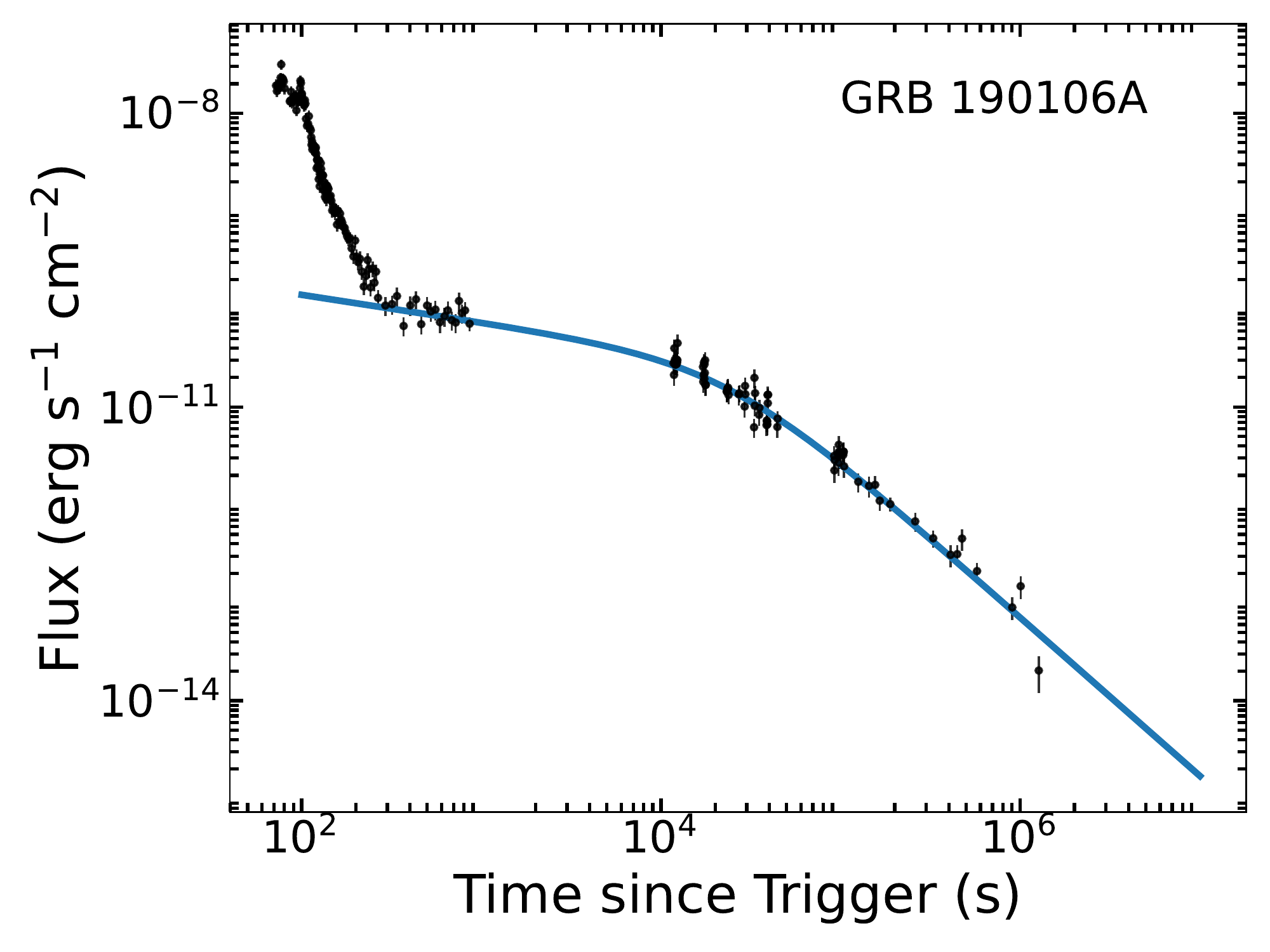}{0.8\textwidth}{}
                   }
\caption{GRB 190106A as a typical event that has a plateau phase
in the X-ray afterglow light curve. The solid curve corresponds to
the best fit result by using Equation (\ref{func1}), through an
MCMC method.
\label{fig1}}
\end{figure}

The Markov Chain Monte Carlo (MCMC) algorithm is employed to fit
the GRB afterglow light curves. The key parameters derived in the
fitting process are presented in Table \ref{tab:1} and Table \ref{tab:2}
for the Starting Sample and the Target Sample, respectively. An example
showing our best fit to the plateau phase and the subsequent X-ray
light curve of GRB 190106A is illustrated in Figure \ref{fig1}.


After obtaining these key parameters, the isotropic luminosity at
the end time of the plateau phase can then be calculated as
\begin{eqnarray} \label{func2}
L_{\rm X} & = & 4 \pi D_{\mathrm{L}}^{2}(z)
\frac{F_{X 0}}{2} K,
\end{eqnarray}
where $D_{\mathrm{L}}(z)$ is the luminosity distance, i.e.,
\begin{eqnarray} \label{func3} D_{\mathrm L}(z) & = &
(1+z)\frac{C}{H_{0}} \int_{0}^{z}\frac{dz^{\prime}}{\sqrt{\Omega
_{M}(1+z^{\prime})^3+\Omega _{\Lambda }} }  ,
\end{eqnarray}
where $z$ is the redshift and $C$ is the speed of light. In this
study, a flat $\Lambda$CDM cosmology with parameters of $H_{0}=$
70.0 $\mathrm{km}$  $\mathrm{s}^{-1}$ $\mathrm{Mpc}^{-1}$, $\Omega
_{M}=0.286$ and $\Omega _{\Lambda}=1-\Omega _{M}$ is adopted.
Meanwhile, the $K$-correction is considered in our calculations,
which is \citep{Bloom2001AJ}
\begin{eqnarray} \label{func4}
K & = & \frac{\int_{E_{1} /(1+z)}^{E_{2} / (1+z)} E N(E) d
E}{\int_{E_{1}} ^{E_{2}} E N(E) d E} ,
\end{eqnarray}
where ($E_{1},E_{2}$) stands for the energy band of the detector.
Equation (4) of \cite{Schaefer2007ApJ} gives a general
form of the GRB photon spectrum.
Considering the relatively narrow energy bandpass of XRT onboard
Swift, we adopt a simple power-law function for the photon
spectrum, e.g., $N(E) = N_{0}E^{-\beta _{X}}$, where $N_{0}$ is a
normalization coefficient and $\beta _{X}$ is the photon index
 measured by Swift XRT. Consequently, the isotropic X-ray
luminosity at the break time is
\begin{eqnarray} \label{func5}
L_{\rm X} & = & \frac{4 \pi D_{\mathrm{L}}^{2}(z)}{(1+z)^{2-\beta_{X}}}
\frac{F_{X 0}}{2},
\end{eqnarray}
after the $K$-correction.

Similarly, the isotropic $\gamma$-ray energy of the prompt phase
(after $K$-correction) is calculated as
\begin{eqnarray} \label{func6}
E_{\gamma, \rm iso} = \frac{4 \pi D_{\mathrm{L}}^{2}(z)S} {(1+z)}
K = \frac{4 \pi D_{\mathrm{L}}^{2}(z) S} {(1+z)^{3-\alpha
_{\gamma}}},
\end{eqnarray}
where $S$ and $\alpha _{\gamma}$ are the fluence and the photon
spectral index measured by Swift/BAT, respectively. The parameters
of $\alpha _{\gamma}$ and $\beta _{X}$ are taken from the Swift
GRB database
\footnote{\url{https://swift.gsfc.nasa.gov/archive/grb_table/}}.

\begin{figure}[ht!]
\gridline{\fig{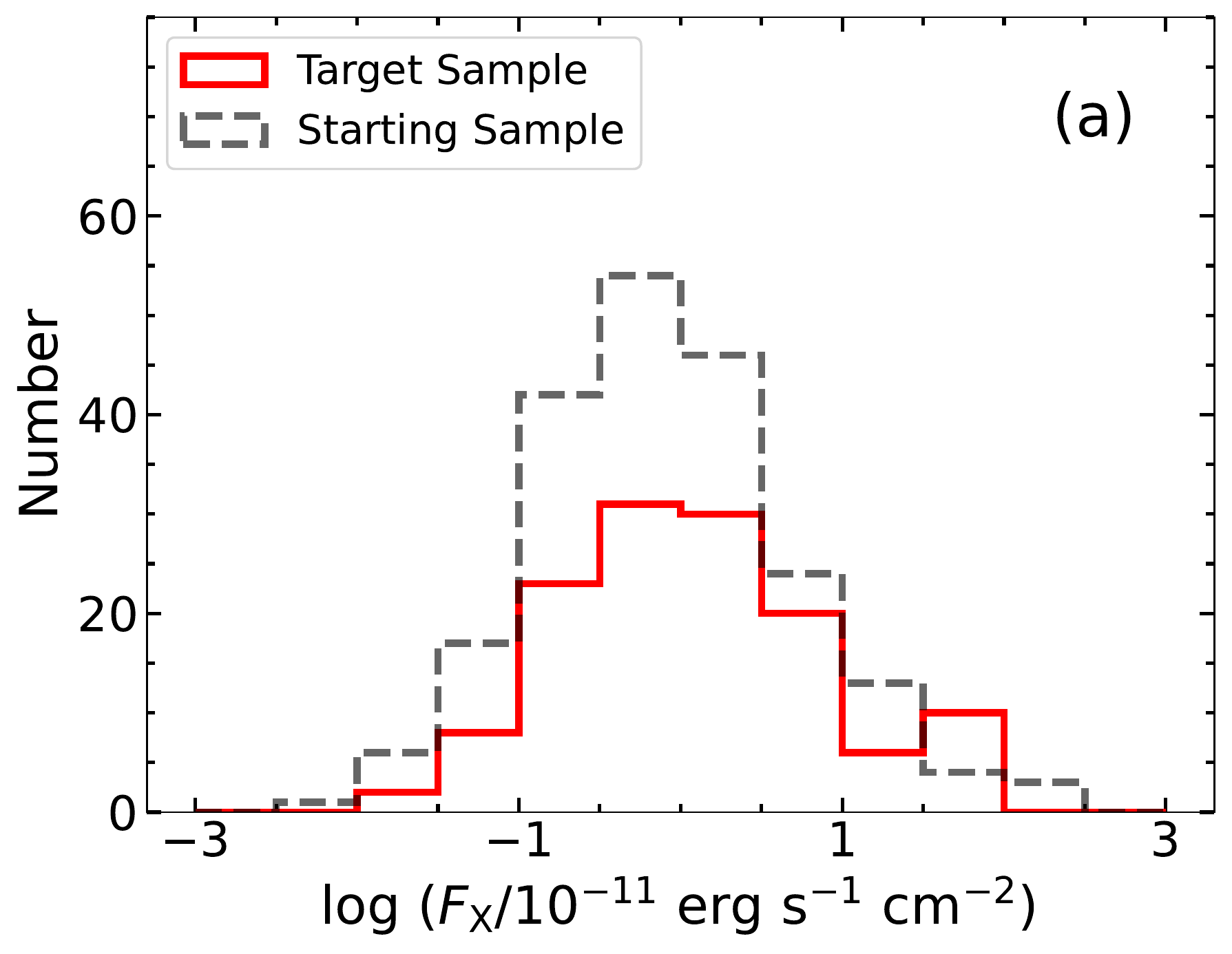}{0.45\textwidth}{}
                   }
\gridline{\fig{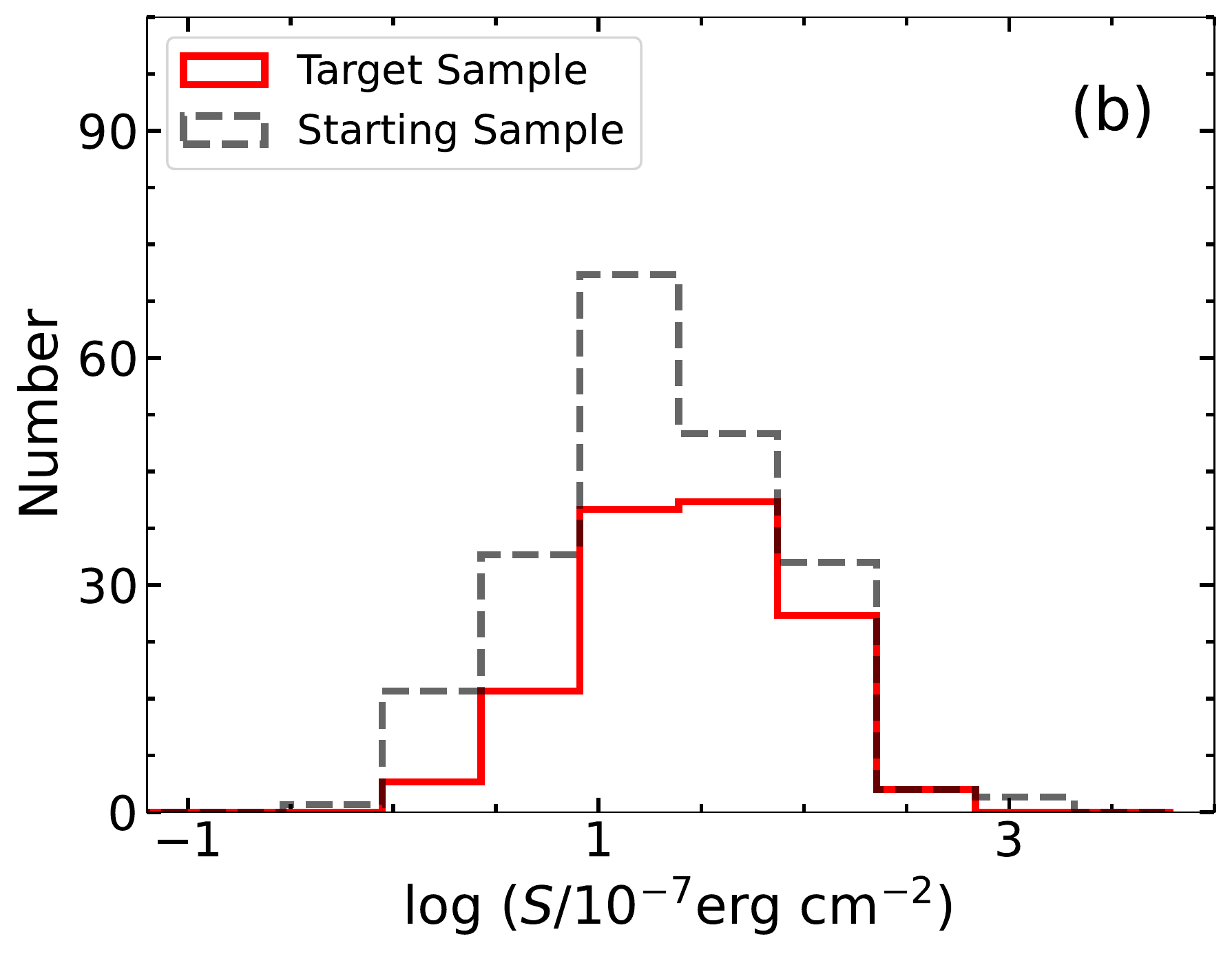}{0.45\textwidth}{}
                   }
\gridline{\fig{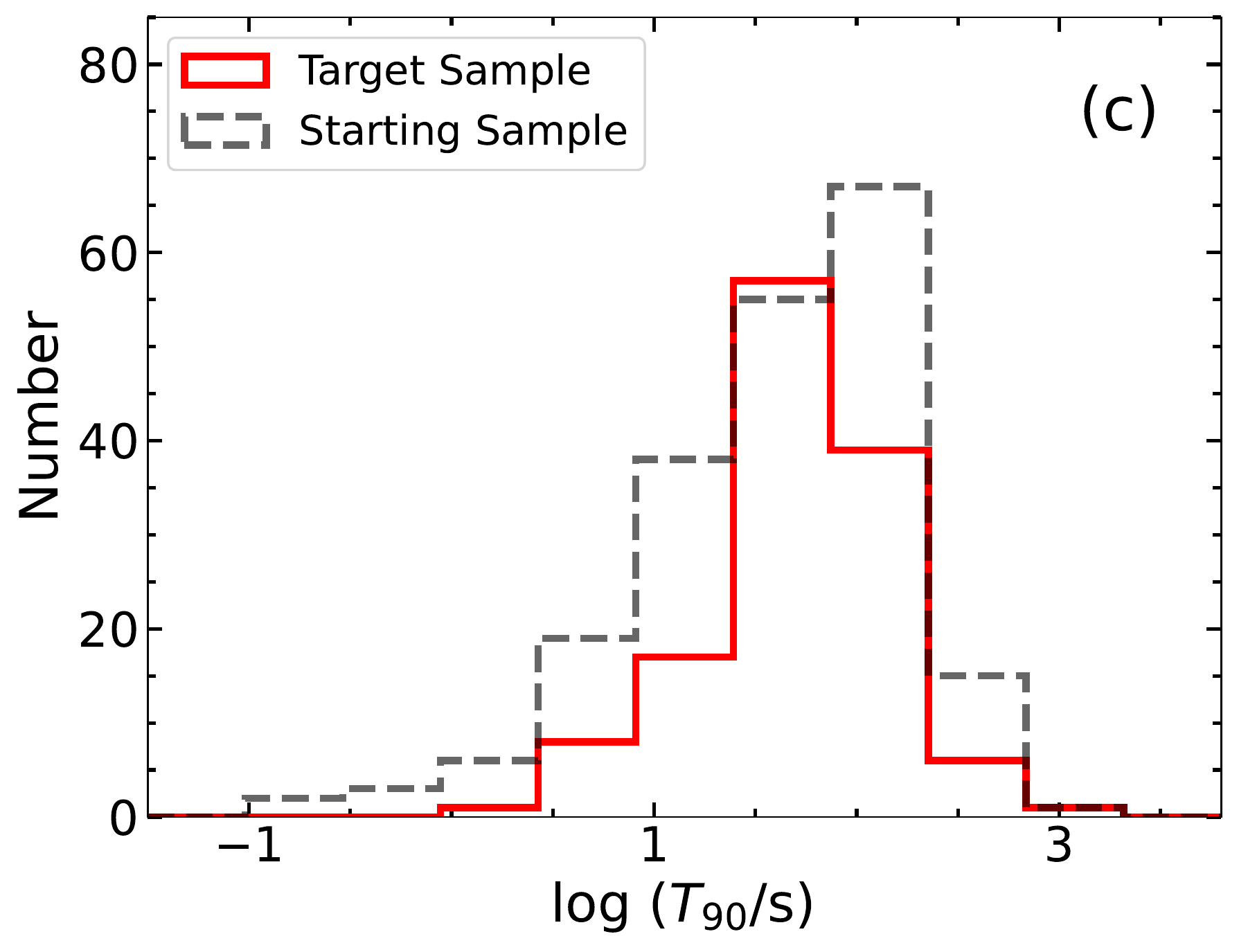}{0.45\textwidth}{}
                   }
\caption{Distribution of $F_{X}$, $S$ and $T_{90}$ for the two
samples. For the Starting Sample (the dashed lines), the mean
values of $F_{X}$, $S$, and $T_{90}$ are $0.8 \times 10^{-11} \rm
erg~cm^{-2}~s^{-1}$, $2.04 \times 10^{-6} \rm erg~cm^{-2}$ and
39.8 s, respectively. Correspondingly, for the Target Sample
(the solid lines), the mean values of $F_{X}$, $S$,
and $T_{90}$ are $1.2 \times 10^{-11} \rm erg~cm^{-2}~s^{-1}$,
$2.69 \times 10^{-6} \rm erg~cm^{-2}$ and 47.9 s, respectively.
\label{fig2}}
\end{figure}

\begin{figure}[ht!]
\gridline{\fig{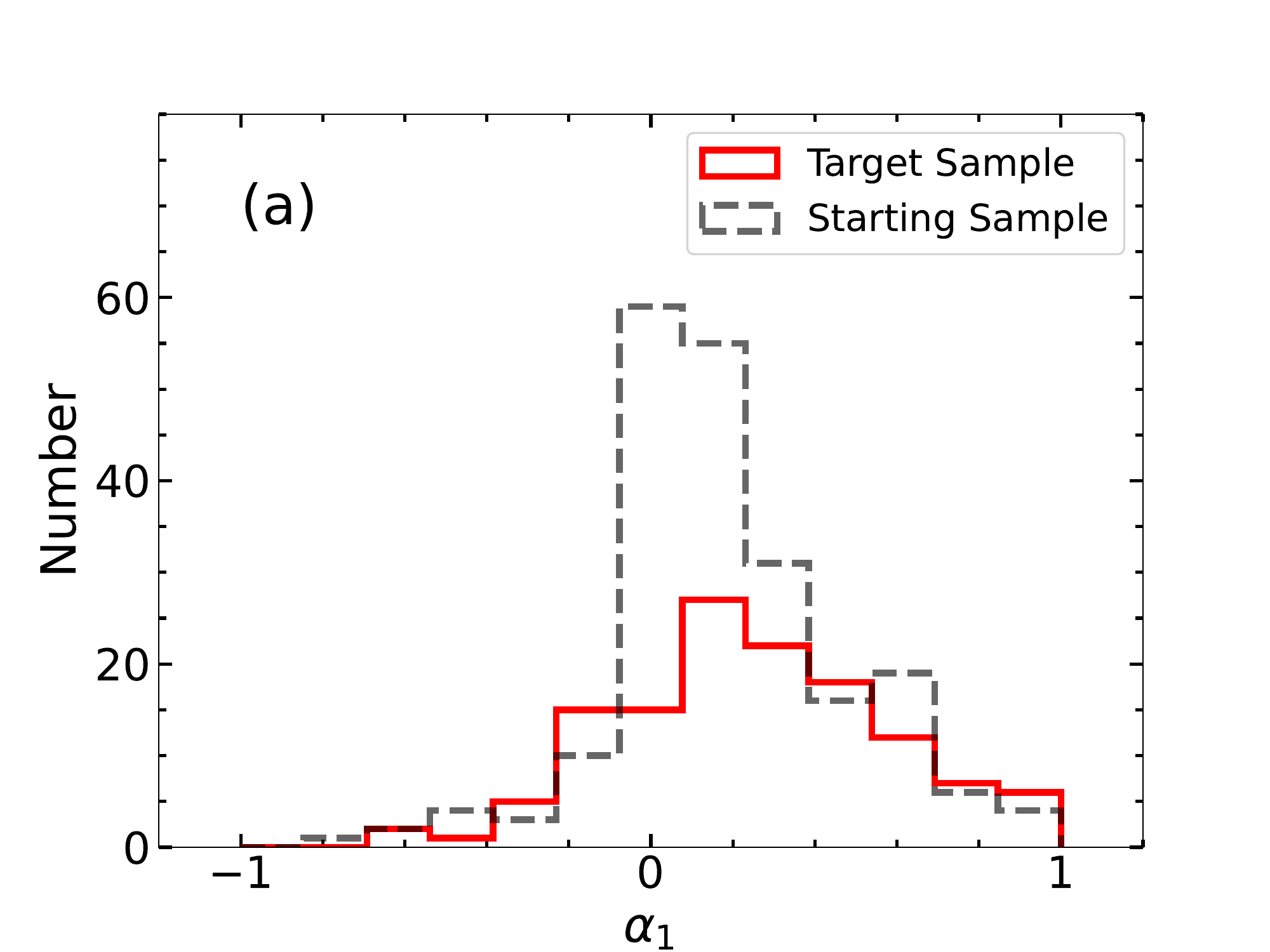}{0.7\textwidth}{}
                   }
\gridline{\fig{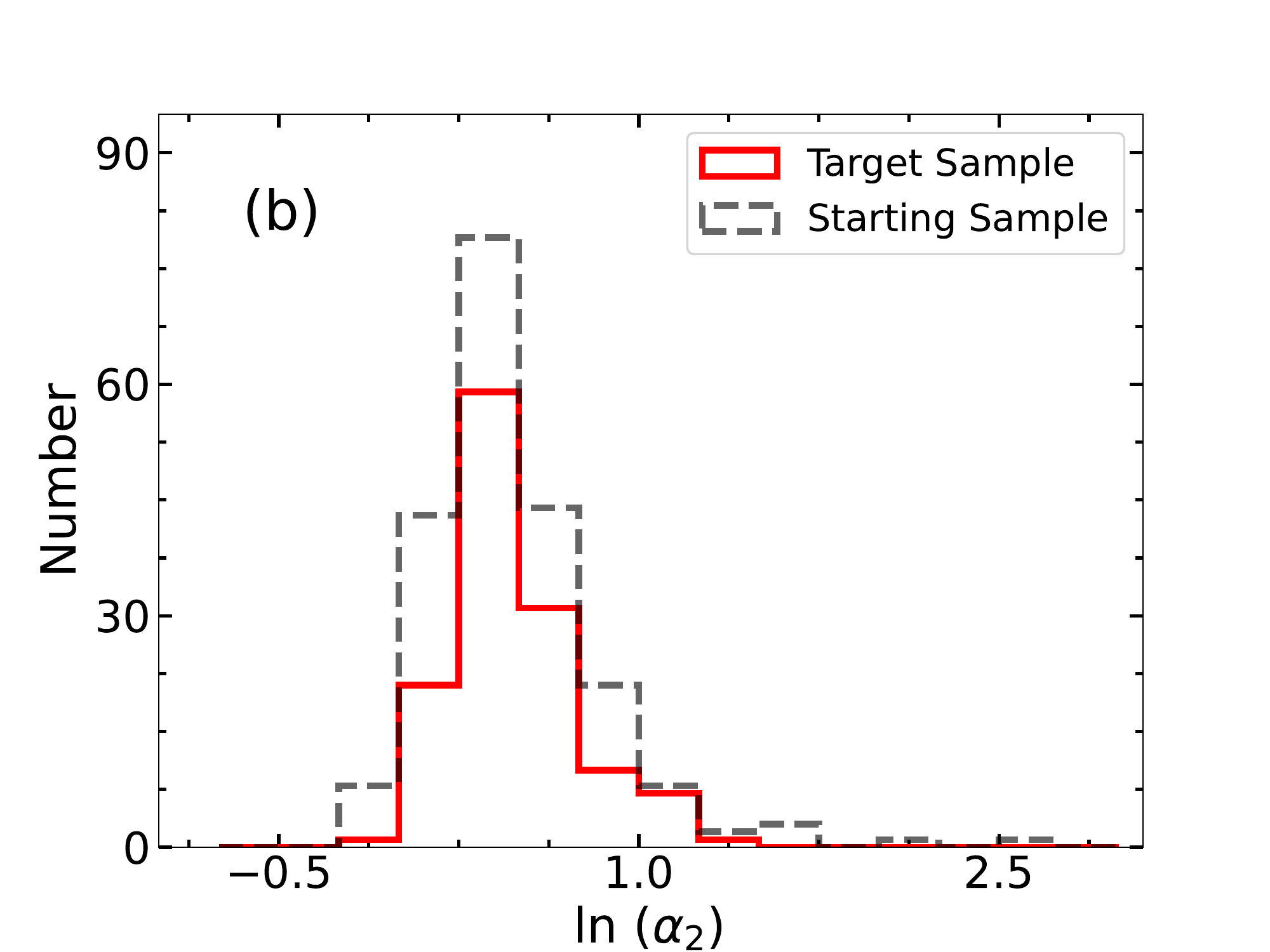}{0.7\textwidth}{}
                   }
\caption{Distribution of the power-law index of the plateau phase
($\alpha_{1}$) and the index after the plateau phase
($\alpha_{2}$). The mean values of $\alpha_{1}$ and $\alpha_{2}$
are 0.19 and 1.77, respective, for the Starting Sample (the dashed
lines). The corresponding mean values are 0.24 and 1.68 for the
Target Sample (the solid lines).
\label{fig3}
}
\end{figure}

A preliminary analyse has been carried out on the basic features of
these two samples. Figure \ref{fig2} presents the distributions of
the plateau flux ($F_{X}$, 0.3 --- 10 keV), the Swift/BAT fluence
($S$, 15 --- 150 keV) and the duration ($T_{90}$). For the Starting
Sample, we see that the mean values of $F_{X}$, $S$, and
$T_{90}$ are $0.8 \times 10^{-11} \rm erg~cm^{-2}~s^{-1}$,
$2.04 \times 10^{-6} \rm erg~cm^{-2}$ and 39.8 s, respectively.
Correspondingly, for the Target Sample, the mean values of $F_{X}$, $S$,
and $T_{90}$ are $1.2 \times 10^{-11} \rm erg~cm^{-2}~s^{-1}$,
$2.69 \times 10^{-6} \rm erg~cm^{-2}$ and 47.9 s, respectively.  We see
that the distributions of these three parameters are largely similar
for the two samples, indicating that there is no systematic difference
between them. The distributions of the temporal power-law decay
indices of the plateau phase ($\alpha_{1}$) and the post-break
segment ($\alpha_{2}$) are shown in Figure \ref{fig3}. The mean
values of $\alpha_{1}$ are 0.19 and 0.24 for the Starting Sample
and the Targeting Sample, respectively. Correspondingly, the mean
values of $\alpha_{2}$ are 1.77 and 1.68 for them. Again, we see
that the distributions of $\alpha_{1}$ and $\alpha_{2}$ are similar
for the two samples.


\section{The $L-T-E$ correlation updated}
\label{sec:LTE}

\citet{Tang2019} derived the $L-T-E$ correlation as $L_{X}\propto
T_{a}^{-1.01}E_{\gamma ,\rm iso}^{0.85}$ with a sample of 174
GRBs. In this section, we utilize our expanded sample, i.e. the
Starting Sample, which includes 210 GRBs, to further study the
$L-T-E$ correlation. To begin with, let us write down the three
parameter correlation in a general form of \citep{xu2012,Tang2019}
\begin{eqnarray} \label{func7}
\log \left(L_{\rm X} / 10^{47} \mathrm{erg}~ \mathrm{s}^{-1}\right) & = & a+b
\log \left(T_{a} / 10^{3} \mathrm{~s}\right)+c \log \left(E_{\gamma,
\rm iso} / 10^{53} \mathrm{erg}\right),
\end{eqnarray}
where, $a$, $b$, and $c$ are coefficients to be determined through
observational data. For each GRB, the parameters of $L_{X}$,
$T_{a}$, and $E_{\gamma, \rm iso}$ can be taken from Table \ref{tab:1}.

An MCMC algorithm is utilized to fit the observational data to
derive the updated $L-T-E$ correlation. To do so, we adopt a joint
likelihood function that is widely used in multi-parameter fitting
process, i.e. \citep{D'Agostini2005}
\begin{eqnarray} \label{func8}
\mathcal{L}\left(a, b, c, \sigma_{\mathrm{int}}\right) \propto \prod_{n}
\frac{1}{\sqrt{\sigma_{\mathrm{int}}^{2}+\sigma_{y,n}^{2}+b^{2} \sigma_
{x_{1, n}}^{2}+c^{2} \sigma_{x_{2, n}}^{2}}}
\exp \left[-\frac{\left(y_{n}-a-b x_{1, n}-c x_{2, n}\right)^{2}}{2
\left(\sigma_{\mathrm{int}}^{2}+\sigma_{y,n}^{2}+b^{2}
\sigma_{x_{1, n}}^{2}+c^{2} \sigma_{x_{2, n}}^{2}\right)}\right],
\end{eqnarray}
where $n$ is the sample size and $\sigma _{\mathrm{int} }$
represents possible intrinsic scatter caused by some unknown
variables. In our calculations, $y$, $x_{1}$ and $x_{2}$ represent
$\log \left(L_{\rm X} / 10^{47} \mathrm{erg}~\mathrm{s}^{-1}\right)$,
$\log \left(T_{a} / 10^{3} \mathrm{~s}\right)$ and $\log
\left(E_{\gamma, \rm iso} / 10^{53} \mathrm{erg}\right)$,
respectively. $\sigma _{y}$, $\sigma _{x_{1}}$ and $\sigma
_{x_{2}}$ are the corresponding error bars.

\begin{figure}[ht!]
\gridline{\fig{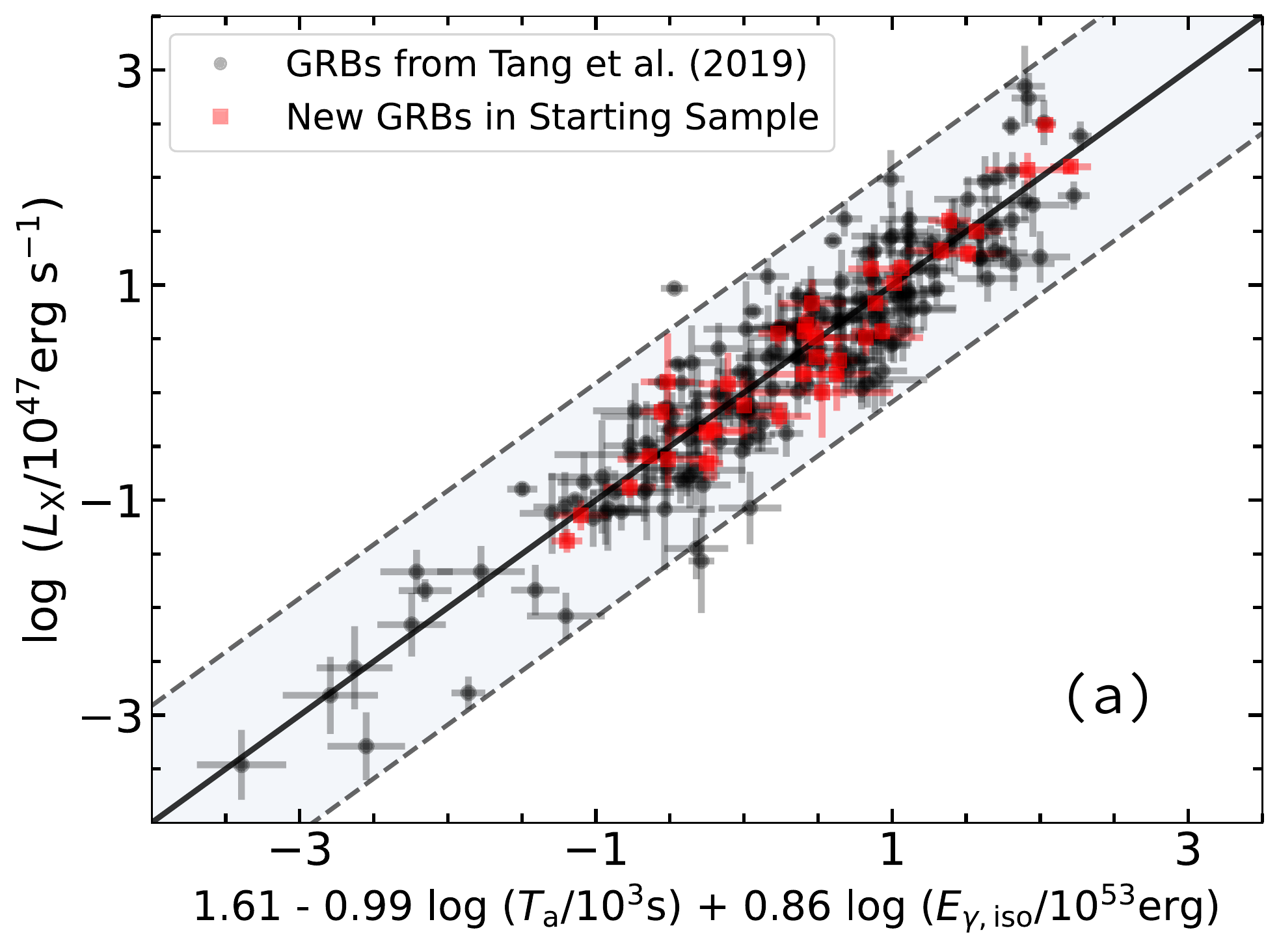}{0.6\textwidth}{}
             }
\gridline{\fig{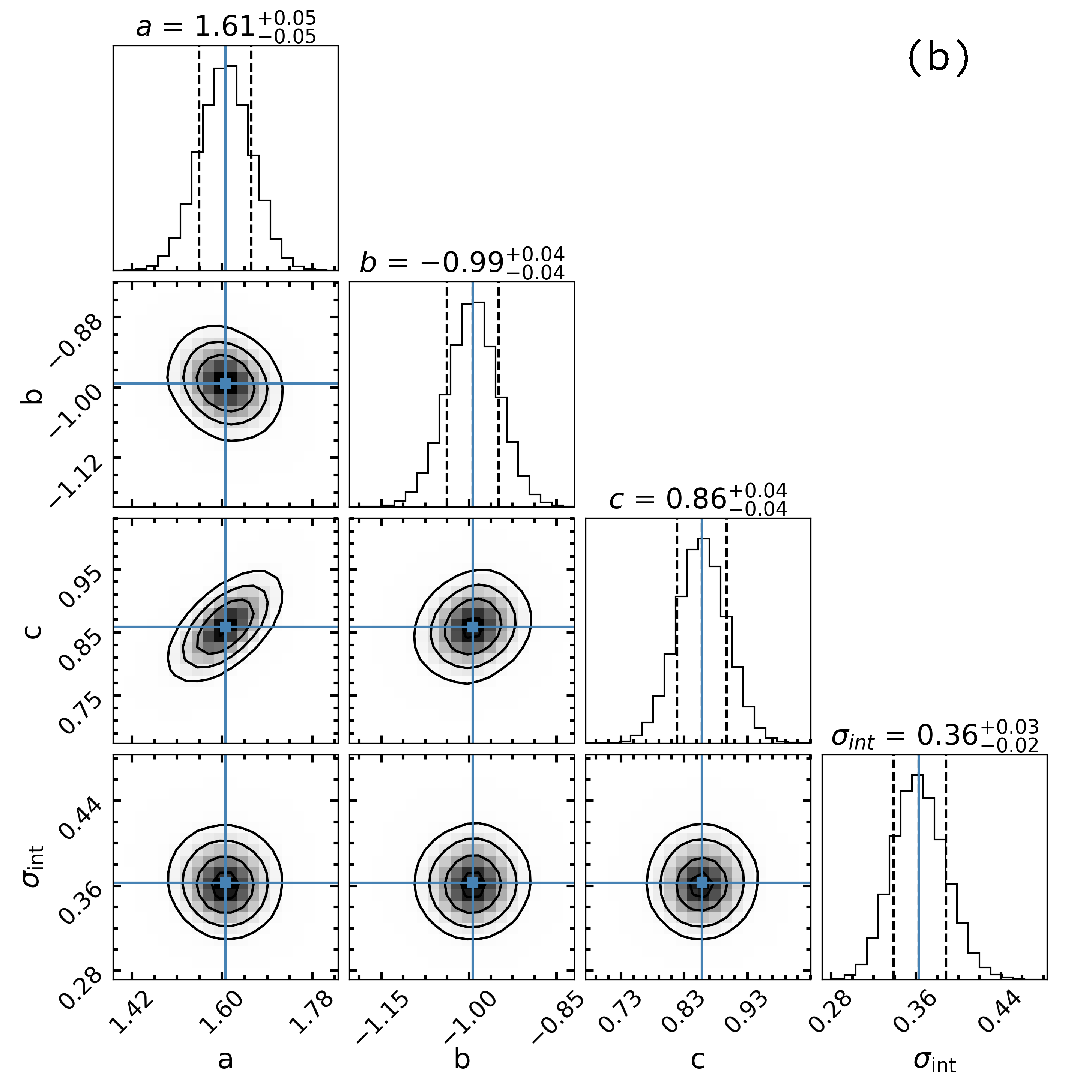}{0.63\textwidth}{}
             }
\caption{(a) The $L-T-E$ correlation updated with the 210 GRBs of
the Starting Sample. The solid line is the best-fit result, and
the dashed lines represent the $3\sigma$ confidence level. (b)
Contour plots showing the uncertainty of the parameters derived
from the MCMC algorithm. The vertical solid lines are the median
values and the vertical dashed lines represent the $1\sigma$
confidence levels.\label{fig4}}
\end{figure}

For the 210 GRBs in the Starting Sample, the best-fit $L-T-E$
correlation is
\begin{eqnarray} \label{func9}
\log \left(L_{\mathrm{X}} / 10^{47} \mathrm{erg}~\mathrm{s}^{-1}
\right) & = & (1.61 \pm 0.05)+(-0.99 \pm 0.04) \times \log
\left(T_{a} / 10^{3} \mathrm{~s}\right)+(0.86 \pm 0.04)\times \log
\left(E_{\gamma, \text {iso}} / 10^{53} \mathrm{erg}\right).
\end{eqnarray}
The fitting results are illustrated in Figure \ref{fig4}. We see
that Equation (\ref{func9}) is well consistent with previous
results of \citet{Tang2019}, who derived the correlation as
$L_{X}\propto T_{a}^{-1.01}E_{\gamma, \rm iso}^{0.85}$, with $\sigma
_{\mathrm{int}} = 0.40 \pm 0.03$. Our study further confirms the
existence of the $L-T-E$ correlation. Note that our intrinsic
scatter parameter is $\sigma _{\mathrm{int}} = 0.36 \pm 0.03$,
which is slightly smaller than that obtained by \citet{Tang2019}.
It means that the correlation is even tighter for this expanded
sample.

\section{Pseudo redshifts from the $L-T-E$ correlation}
\label{sec:estimator}

The three-parameter $L-T-E$ correlation connects redshift with the
observed quantities of $S$ (the fluence), $T_{a}$, and $F_{X}$
(the X-ray flux). It hints us that we could employ the correlation
to estimate the redshift for GRBs in the Target Sample, whose
redshifts are all unknown. For this purpose, we first define a
function $g(z)$ as
\begin{eqnarray} \label{func10}
\begin{aligned}
g(z) =& \log \left(L_{\rm X} / 10^{47} \mathrm{erg}  \mathrm{~s}^{-1}\right)
- a-b\log \left(T_{\rm a} / 10^{3} \mathrm{~s}\right)-c \log \left(E_{\gamma,
\text{iso}} / 10^{53} \mathrm{erg}\right) \\
=& (2-2 c) \log \left(\int_{0}^{z}
\frac{1}{\sqrt{\left(1+z^{\prime}\right)^{3}
\Omega_{M}+\Omega_{\Lambda}}} d
z^{\prime}\right)+\left(b+c+\beta_{X}- c
\alpha_{\gamma}\right) \log \left(1+z \right)  \\-&c \log\left(4 \pi
S\right)+\log\left(2 \pi F_{X0}\right)-a-b \log
\left(T_{a,\text{obs}}\right)+3 b-47+53 c +(2-2 c) \log
\left(\frac{C}{H_{0}}\right).
\end{aligned}
\end{eqnarray}
In our calculations, the best fitting values
derived in Section \ref{sec:LTE} are adopted for the coefficients
$a$, $b$ and $c$, e.g. $a = 1.61$, $b = -0.99$ and $c = 0.86$.

\begin{figure}[ht!]
\gridline{\fig{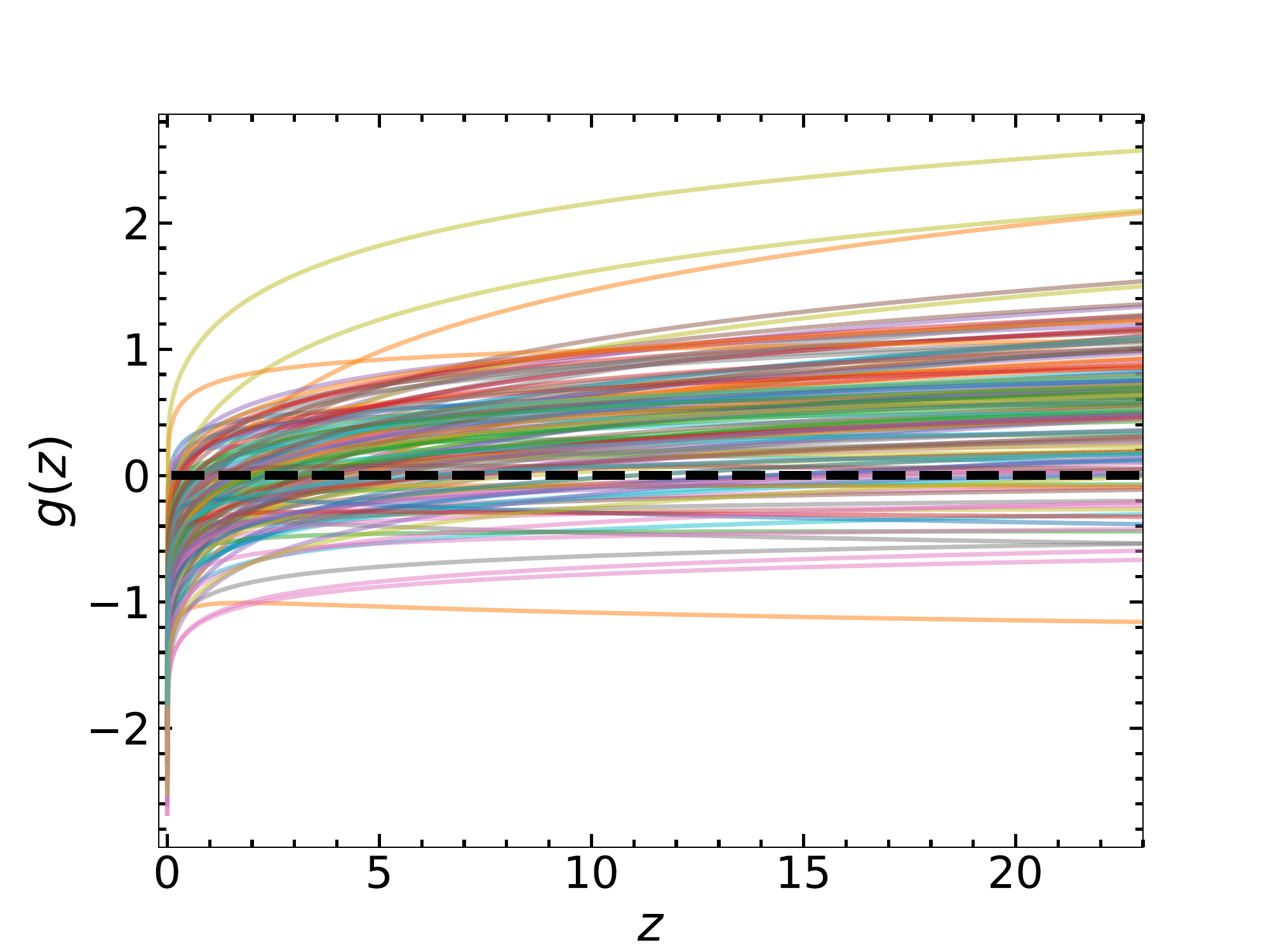}{0.8\textwidth}{}
                   }
\caption{The function $g(z)$ as defined in Equation (\ref{func10})
for all the 130 GRBs in the Target Sample. The pseudo redshift of
each GRB is determined by the intersection of the $g(z)$ curve
with the horizontal dashed line ($g = 0$). For those GRBs whose
$g(z)$ curve does not intersect with the dashed line, no pseudo
redshift can be derived. \label{fig5}}
\end{figure}

If a particular GRB strictly satisfies the $L-T-E$ correlation,
then the corresponding $g(z)$ should be zero. So we can
calculate the pseudo redshift of a GRB by solving the equation of
$g(z)=0$ when the $F_{X}$, $S$, $T_{\rm a,obs}$,
$\alpha_{\gamma}$, $\beta_{X}$ parameters are available. In other
words, the pseudo redshift is the intersection of the function
$g(z)$ with the horizontal axis in the $g(z)-z$ plane. As
a demonstration, the function $g(z)$ is plot for all target
GRBs in Figure \ref{fig5}. We can see that the $g(z)$ curve is
generally very flat at high redshift. In fact, if the intersection
is at a point with $z > 20$, the derived pseudo redshift will be
highly uncertain due to the existence of the error bars of the
input parameters, i.e. $F_{X}$, $S$, $T_{\rm a,obs}$,
$\alpha_{\gamma}$, and $\beta _{X}$. So, in our calculations, we
will limit the valid range of the pseudo redshift in $0 < z < 20$.
It should also be noted that for a small fraction ($\sim 1/6$) of
the target GRBs, the $g(z)$ curve is well under the zero line
up to $z = 20$. It means that these GRBs may have a redshift
larger than 20. However, the possibility that they actually do not
follow the $L-T-E$ correlation cannot be excluded.

Using the three-parameter $L-T-E$ correlation, we have tried to
calculate the pseudo redshifts for GRBs in the Target Sample. Due
to the reason mentioned just above, the redshift can be derived
only for 108 target GRBs. For the remaining 22 GRBs in the sample,
the pseudo redshift is not available.

\begin{figure}[ht!]
\gridline{\fig{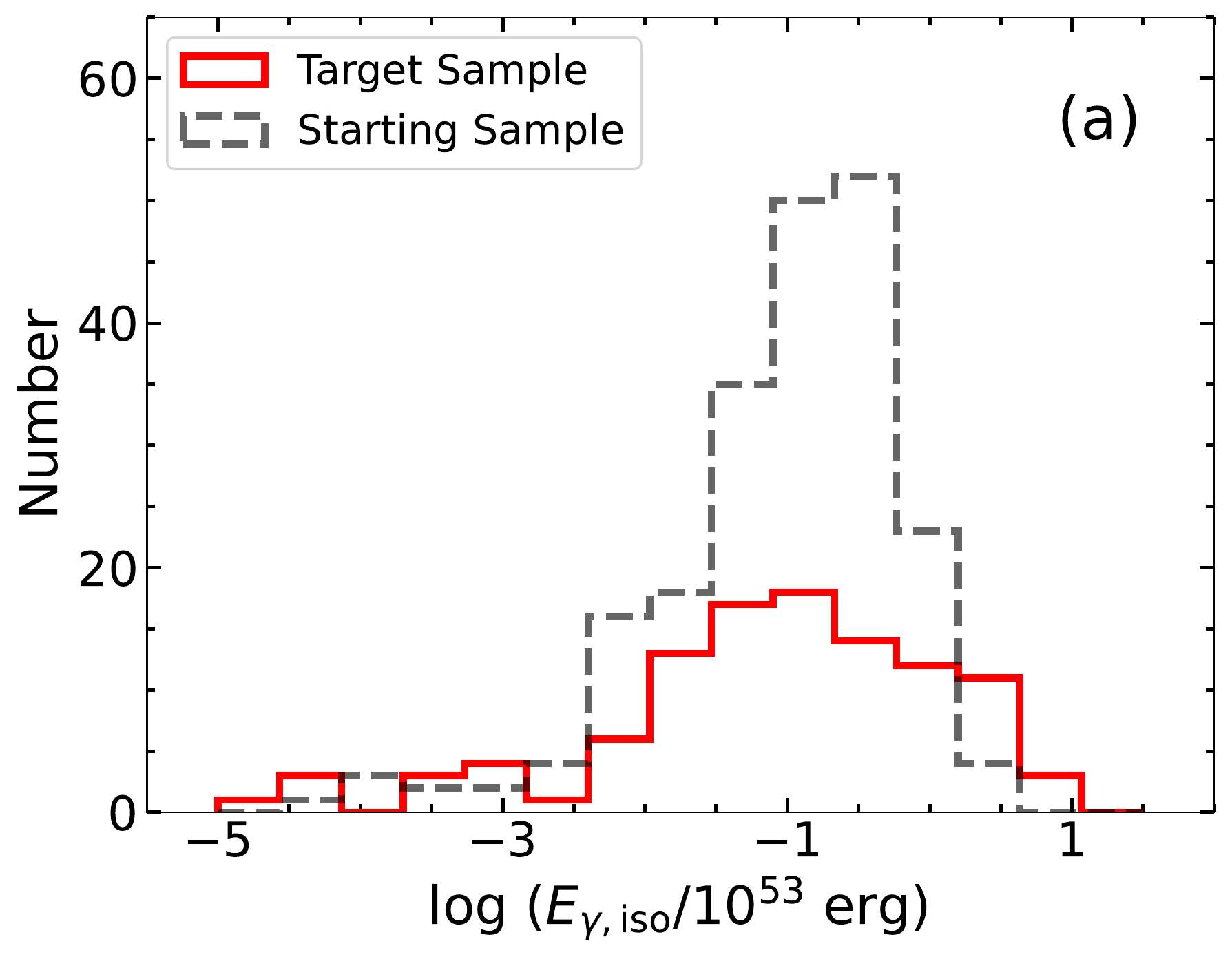}{0.45\textwidth}{}
          \fig{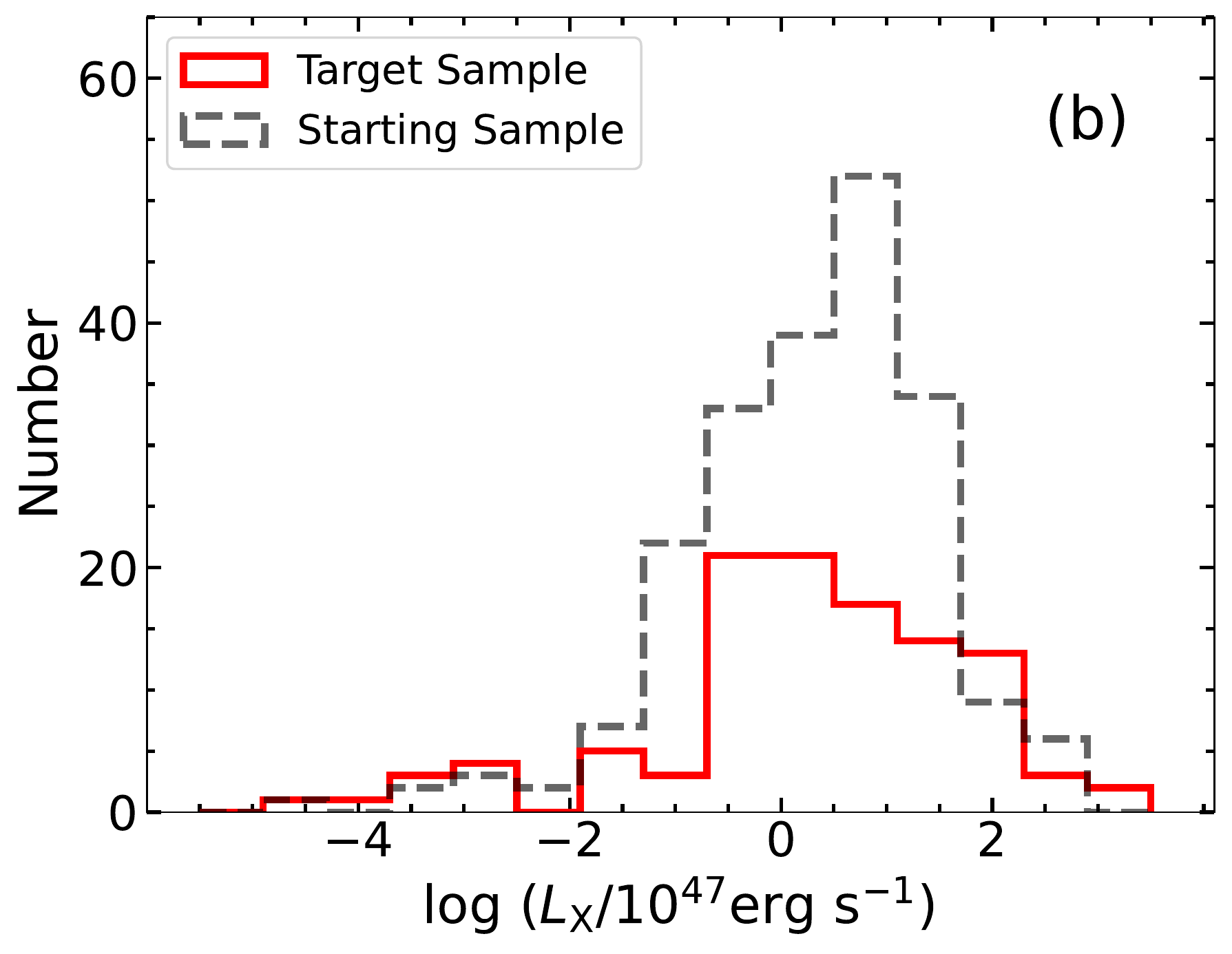}{0.45\textwidth}{}
                   }
\gridline{\fig{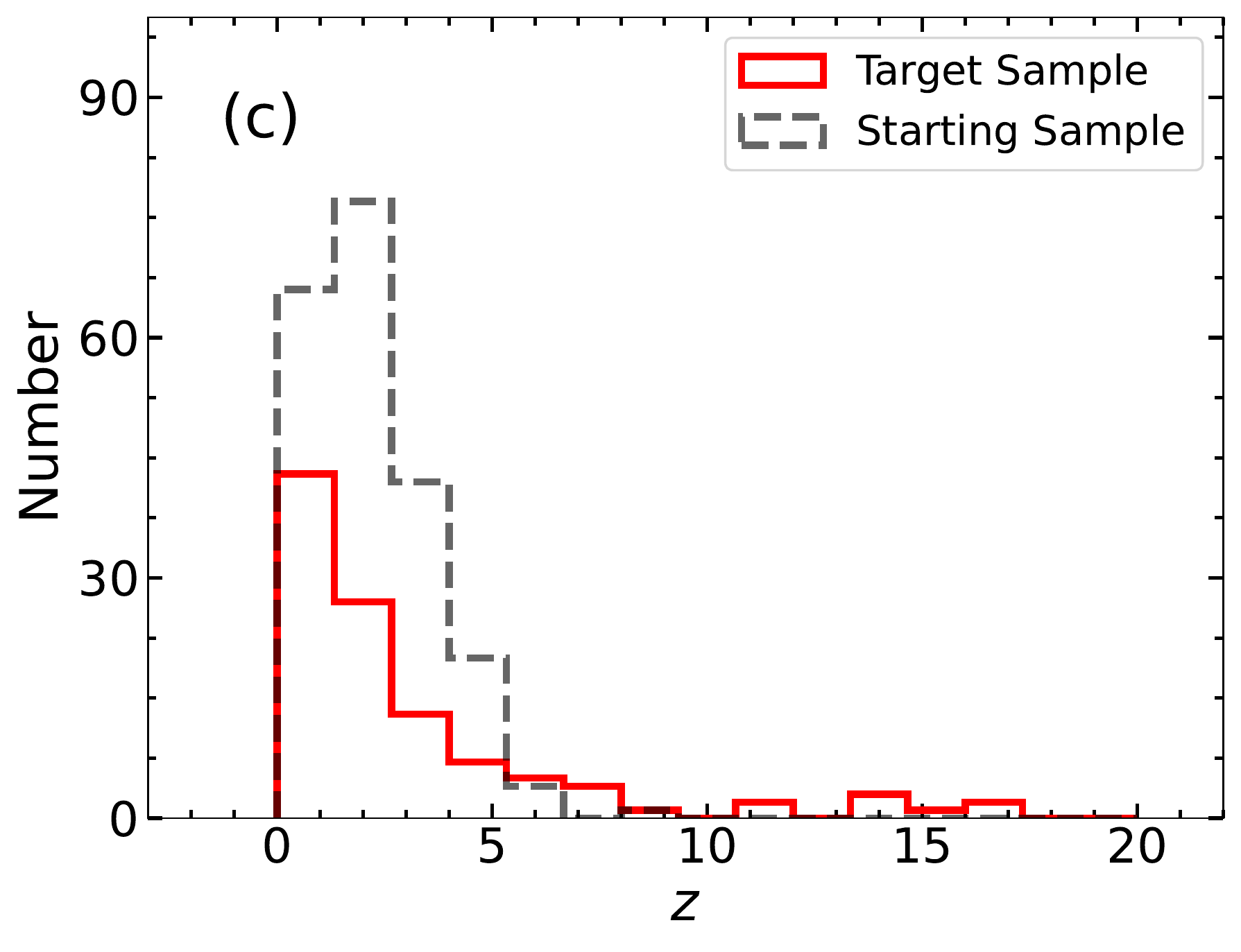}{0.45\textwidth}{}
          \fig{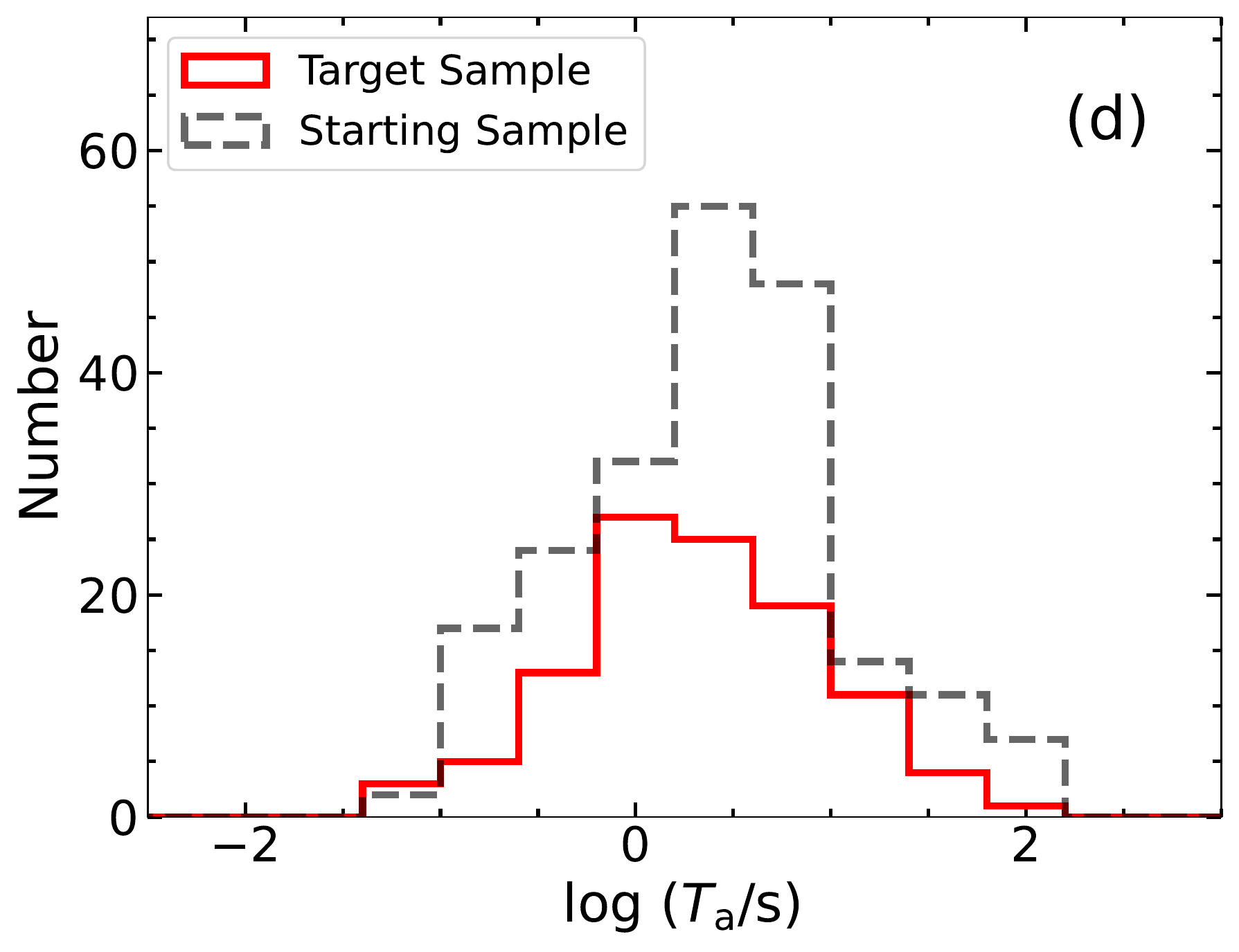}{0.45\textwidth}{}
                   }
\caption{Distribution of $E_{\gamma ,\rm iso}$, $L_{\rm X}$, $z$, and
$T_{\rm a}$ for the two samples. For the Target Sample (the solid lines),
the mean values of $E_{\gamma ,\rm iso}$, $L_{\rm X}$, $z$ and $T_{\rm a}$
are $6.46 \times 10^{51}\rm erg$, $1.82 \times 10^{47}\rm erg~s^{-1}$,
3.08 and $2.14 \times 10^{3}\rm s$, respectively. For the Starting
Sample (the dashed lines), the corresponding mean values are $8.91
\times 10^{51}\rm erg$, $1.91 \times 10^{47}\rm erg~s^{-1}$, 2.2
and $2.51 \times 10^{3}\rm s$, respectively.
\label{fig6}}
\end{figure}

With the pseudo redshift, we can then calculate some key
parameters of the target GRBs, such as $E_{\gamma ,\rm iso}$,
$L_{\rm X}$, and $T_{\rm a}$ (note that this break time is defined in the
GRB rest frame), which are all distance-dependent. Figure
\ref{fig6} shows the distributions of these derived parameters,
together with that of $z$. For the Target Sample, the mean values
of $E_{\gamma ,\rm iso}$, $L_{\rm X}$, $z$ and $T_{\rm a}$ are $6.46
\times 10^{51}\rm erg$, $1.82 \times 10^{47}\rm erg~s^{-1}$, 3.08
and $2.14 \times 10^{3}\rm s$, respectively. As a comparison, for
the Starting Sample, the corresponding mean values are $8.91
\times 10^{51}\rm erg$, $1.91 \times 10^{47}\rm erg~s^{-1}$, 2.2
and $2.51 \times 10^{3}\rm s$, respectively. We see that the
distributions of these parameters are quite similar for the two
samples. The largest difference is observed in the distribution of
$z$. The mean distance of Target Sample is obviously higher than
that of the Starting Sample. This is mainly due to the fact that a
few target GRBs have a very high pseudo redshift, larger than 10.
Whether these events really reside at such high redshifts is an
interesting question.

\begin{figure}[ht!]
\gridline{\fig{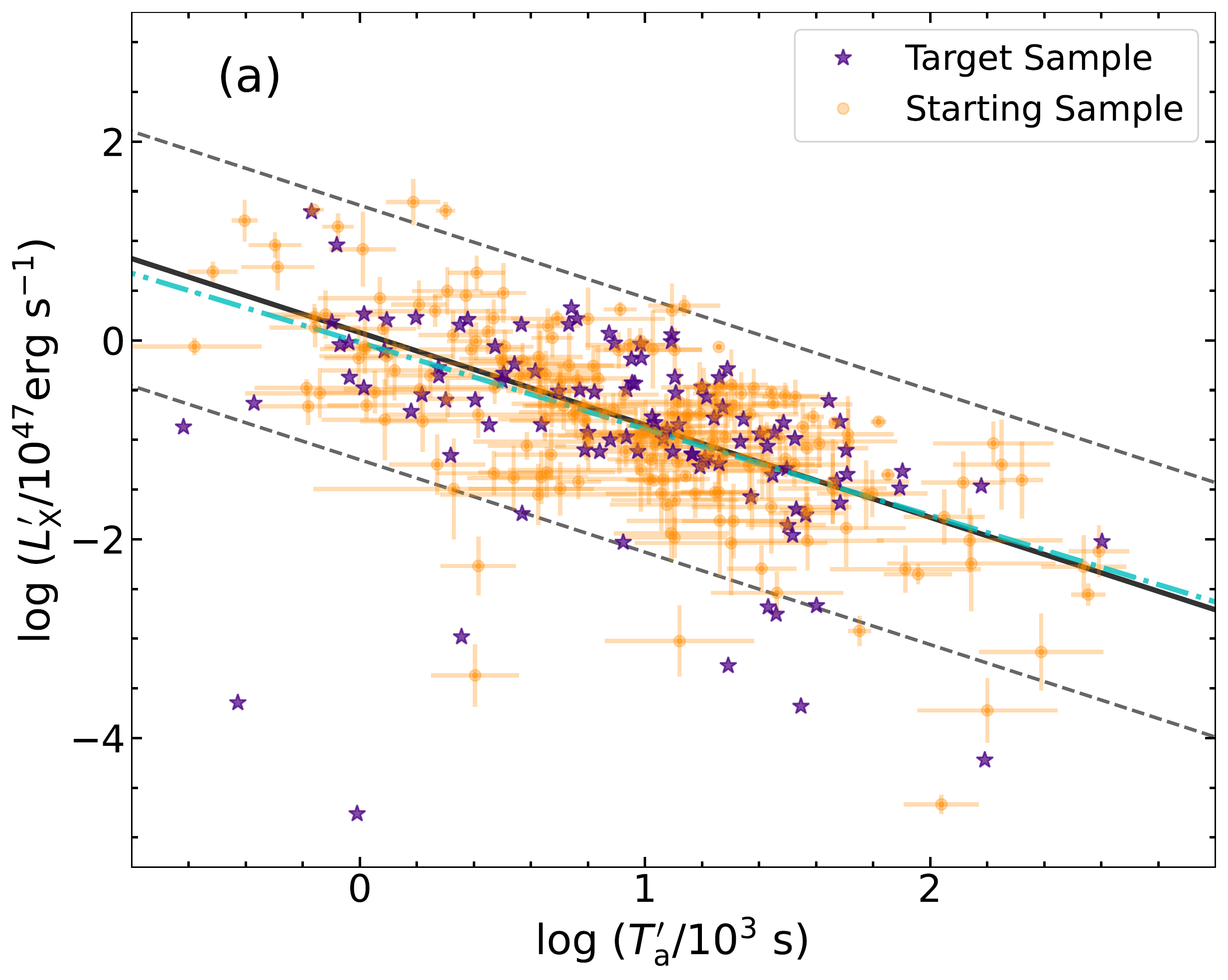}{0.65\textwidth}{}
                   }
\gridline{\fig{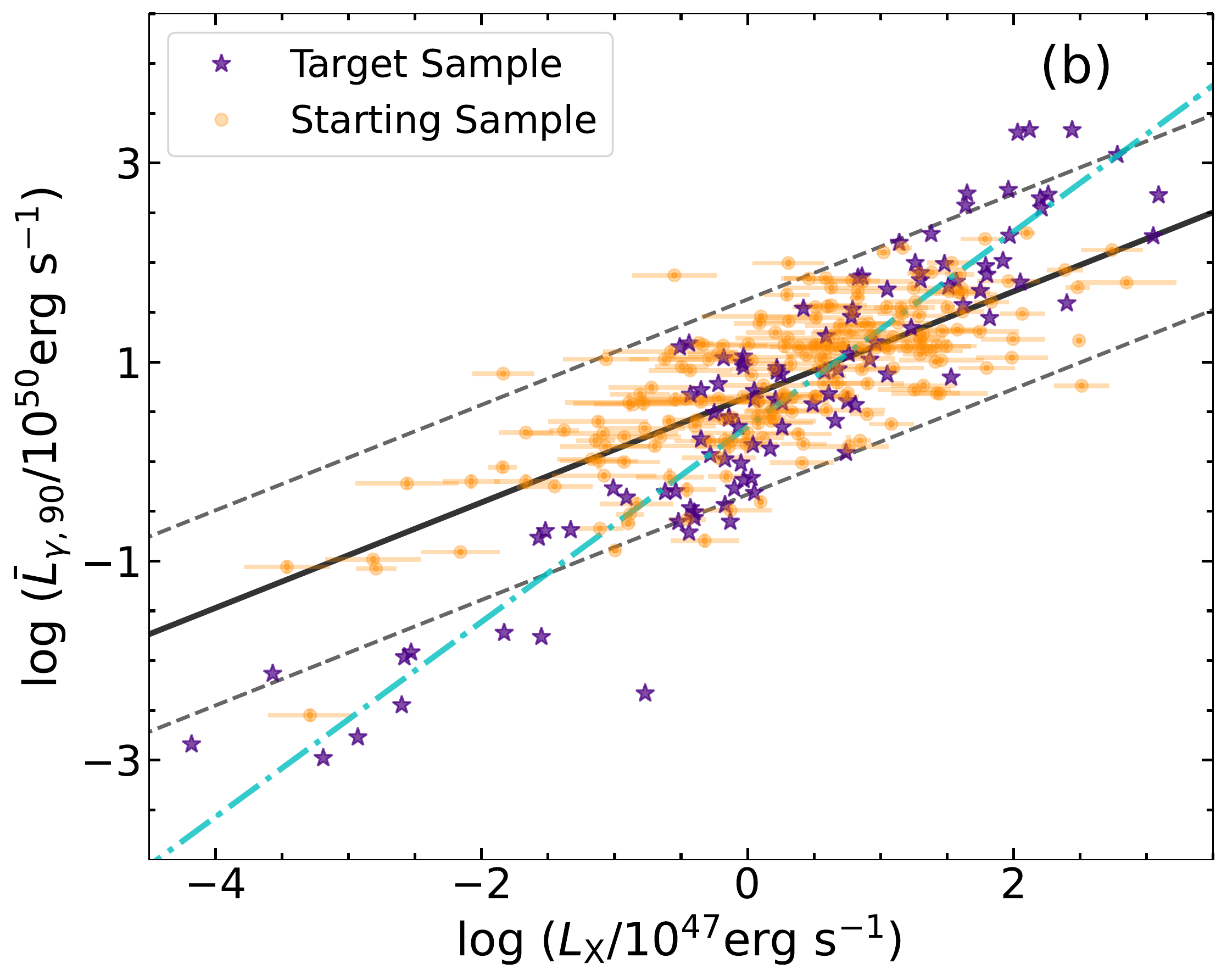}{0.65\textwidth}{}
                   }
\caption{(a) $L^{\prime}_{\rm X}$ plotted versus $T^{\prime}_{\rm a}$
for the two GRB samples. The dot points correspond to the Starting
Sample, which can be best fit as $L^{\prime}_{\rm X} \propto
T_{\rm a}^{\prime -0.93 \pm 0.08}$ (the solid line). The dashed
lines correspondingly show the $2\sigma$ range. The star points
correspond to the Target Sample, which mainly fall within
the above $2\sigma$ range. They can be best fit as
$L^{\prime}_{\rm X} \propto T_{\rm a}^{\prime -0.87 \pm 0.14}$
(the dash-dotted line). Note that the two points at the lower left corner
seem to be obvious outliers and are not included in the fit. 
(b) $\bar L_{\gamma,90}$ plotted versus $L_{\rm X}$ for the two GRB
samples. The dot points correspond to the Starting Sample, which
can be best fit as $\bar L_{\gamma,90} \propto L_{\rm X}^{0.53 \pm
0.03}$ (the solid line). The dashed lines correspondingly show the
$2\sigma$ range. The star points correspond to the Target Sample,
which can be best fit as $\bar L_{\gamma,90} \propto L_{\rm X}^{0.98
\pm 0.04}$ (the dash-dotted line). Note that the $\bar L_
{\gamma,90} - L_{\rm X}$ correlation seems to be different for the
two samples.
\label{fig7}}
\end{figure}

If the pseudo redshift is really a credible measure of the
distance of the GRBs, then it is interesting to examine whether
the Target Sample also satisfy other popular correlations of GRBs.
Let us first take the $L_{\rm X}-T_{\rm a}$ correlation as an
example. \cite{dainotti2008MNRAS} discovered for the first time
that there is an anti-correlation between $L_{\rm X}$ and $T_{\rm
a}$. The effects of selection bias and redshift evolution
on the $L_{\rm X}-T_{\rm a}$ correlation have been investigated in
previous studies \citep{Dainotti2013aApJ,Dainotti2015bMNRAS,
Dainotti2017aA&A,Dainotti2022MNRAS}. Here
we take the power-law evolutionary functions proposed by
\cite{Dainotti2022MNRAS}, e.g., $G(z)=(1+z)^{k_{L_{\rm X}}}$ and
$F(z)=(1+z)^{k_{T_{\rm a}}}$. The local variables are defined as
$L^{\prime}_{\rm X} \equiv L_{\rm X} / G(z)$ and $T^{\prime}_{\rm
a} \equiv T_{\rm a} / F(z)$. The power-law indices, $k_{L_{\rm
X}}$ and $k_{T_{\rm a}}$, are $2.42 \pm 0.58$ and
$-1.25 \pm 0.28$. In Figure \ref{fig7}(a),
$L^{\prime}_{\rm X}$ is plot against $T^{\prime}_{\rm a}$ for the
Target Sample, and the results are compared with that of normal
GRBs with a measured redshift. For normal GRBs, the best fit
$L^{\prime}_{\rm X}-T^{\prime}_{\rm a}$ relation is
$L^{\prime}_{\rm X} \propto T_{\rm a}^{\prime -0.93 \pm 0.08}$, and
it is $L^{\prime}_{\rm X} \propto T_{\rm a}^{\prime -0.87 \pm
0.14}$ for the Target Sample. These results are consistent with
\cite{Dainotti2017aA&A}. We see that these two groups are
well consistent with each other. In fact, most of the target GRBs
fall within the $2\sigma_{\rm int}$ range of normal GRBs.

Another interesting example is the prompt-afterglow
$L_{\rm X}-\bar L_{\gamma,90}$ correlation \citep{Dainotti2011MNRAS,Dainotti2015bMNRAS}. 
Here $\bar L_{\gamma,90}$ is defined as
$\bar L_{\gamma,90} \equiv  E_{\gamma, \rm iso}(1+z)/T_{90}$, which
reflects the average intrinsic luminosity of the prompt emission
phase. Figure \ref{fig7}(b) plots $L_{\rm X}$ versus $\bar L_{\gamma,90}$
for the two GRB groups. We see that there is an obvious correlation
between $L_{\rm X}$ and $\bar L_{\gamma,90}$ for each group. However,
the best fit result is $\bar L_{\gamma,90} \propto L_{\rm X}^{0.98
\pm 0.04}$ for the Target Sample, while it is $\bar
 L_{\gamma,90} \propto L_{\rm X}^{0.53 \pm 0.03}$ for normal GRBs.
There is a clear difference in the power-law index. The reason
of this difference is still unknown and deserves further studying
in the future.

\begin{figure}[ht!]
\gridline{\fig{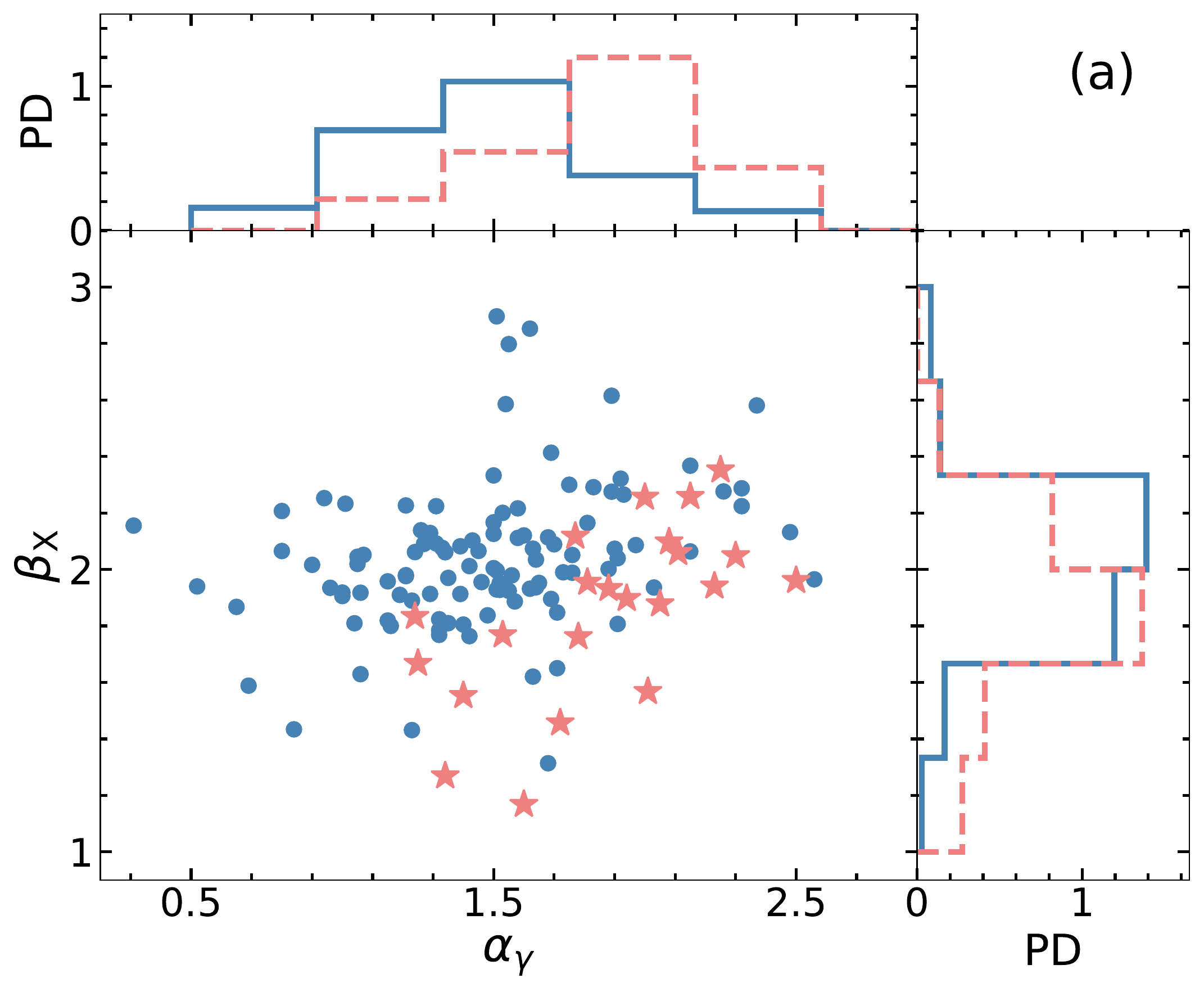}{0.6\textwidth}{}
                   }
\gridline{\fig{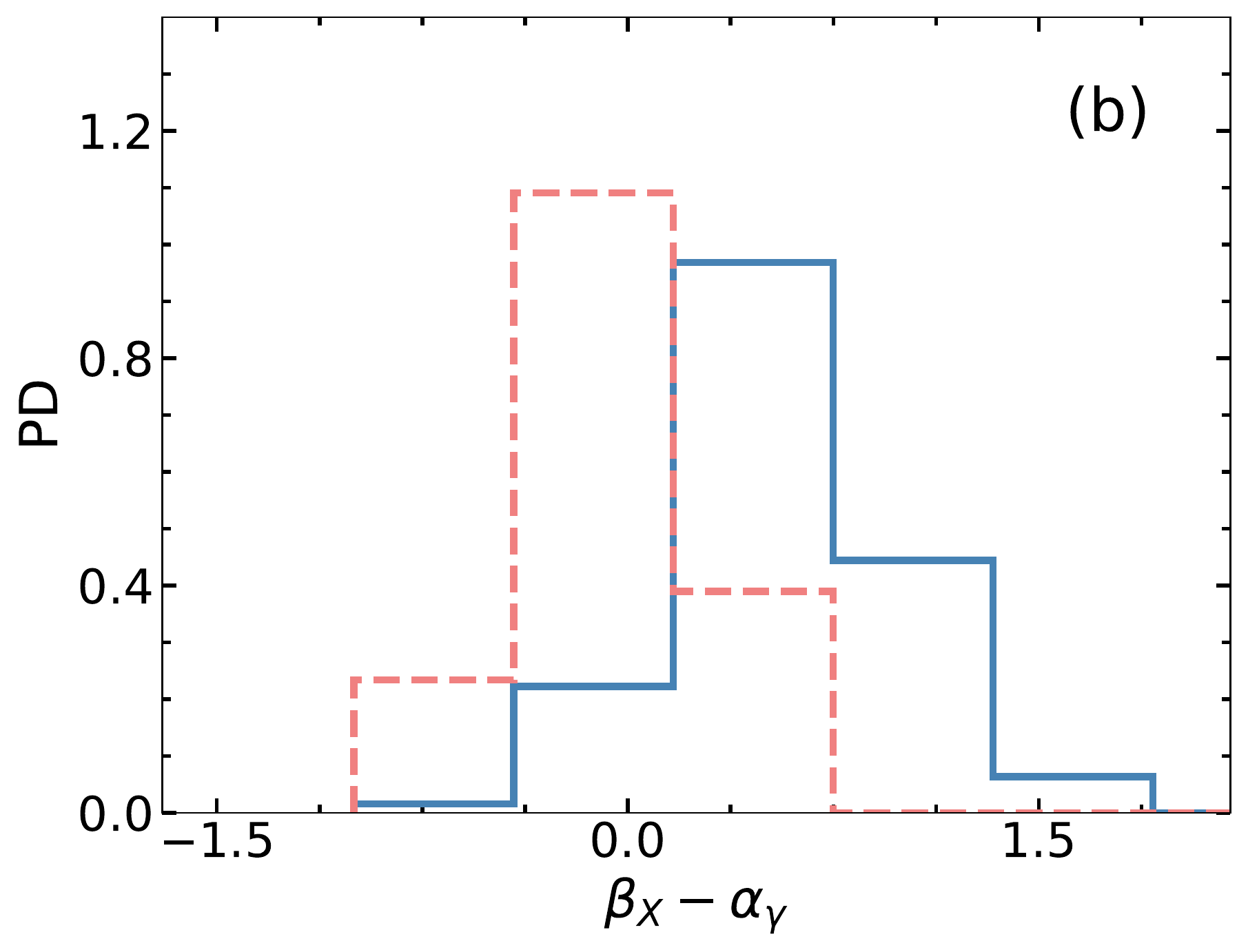}{0.6\textwidth}{}
                   }
\caption{(a) The X-ray photon index ($\beta_{\rm X}$) plotted
versus the BAT $\gamma$-ray photon index ($\alpha_{\gamma}$) for
all the GRBs in the Target Sample. The dot points are the 108
events whose pseudo redshift can be successfully derived from the
$L-T-E$ correlation, while the star symbols correspond to the
remaining 22 events whose pseudo redshift cannot be calculated.
The probability density (PD) distribution of these two groups are
also shown. (b) The distribution of $\beta_{\rm X} -
\alpha_{\gamma}$ for the above two groups. The solid line
corresponds to the 108 GRBs, while the dashed line corresponds to
the remaining 22 GRBs. \label{fig8}}
\end{figure}

As mentioned above, the pseudo redshift could not be derived for
22 GRBs in the Target Sample. Here we present some detailed
discussion on this issue. In Equation (\ref{func10}), we notice
that the coefficient of the item $\log (1+z)$ is
$b + c + \beta_{X} - \mathrm{c} \alpha_{\gamma}$. Since $b \sim -1$
and $c \sim 1$ in our cases, the exact value of this coefficient
is mainly determined by $\beta_{X} - \alpha_{\gamma}$.
If $\beta_{X} \approx \alpha_{\gamma}$, then there will be a
singular point in the function so that the pseudo redshift
could not be solved. To further examine this issue, the photon
indices of the Target Sample are plot in Figure \ref{fig8},
where the 22 GRBs are specially marked by star symbols. In
Figure \ref{fig8}(a), we see that the distribution of these
22 events does deviate from the other GRBs systematically.
Especially, Figure \ref{fig8}(b) clearly shows that the
mean value of $\beta_{X} - \alpha_{\gamma}$ is very close
to zero for these 22 GRBs. This is exactly the reason that their
pseudo redshift cannot be derived.

\section{summary and discussion}
\label{sec:summary}

Many GRBs are characterized by a plateau phase in the X-ray
afterglow light curve. In this study, we have collected a large
sample containing 210 such GRBs with known redshifts. With this
Starting Sample, the three parameter $L-T-E$ correlation is
updated as $L_{\rm X} \propto T_{\rm a}^{-0.99 \pm 0.04}E_{\gamma,\rm
iso}^{0.86\pm0.04}$, which is consistent with the results of
\citet{xu2012} and \cite{Tang2019}. We have also collected another sample,
i.e. the Target Sample, which contains 130 plateau GRBs whose
redshifts were not measured. It is found that the
distributions of many key parameters are similar for the two
samples. The upgraded $L-T-E$ correlation is then applied to the
Target Sample. Pseudo redshifts are calculated for 108 GRBs in
this sample. But for the remaining 22 GRBs, the pseudo redshift
cannot be credibly derived. The reason may be that they are either
at extremely high redshifts or they do not strictly follow the
$L-T-E$ correlation. It could also be due to the fact that they
have an equal photon index in X-rays and $\gamma$-rays, i.e.
$\alpha_{\gamma} \approx \beta_{X}$.

The origin of the plateau phase in the X-ray afterglow of GRBs is
still quite uncertain. \citet{Suvorov2021} argued that it could
not be produced by the fall-back accretion of a black hole because
of the short surviving time-scale ($\sim$ seconds). In fact, the
most popular interpretation is that it is due to the energy
injection from a millisecond magnetar, through dipolar radiation
\citep{Duncan1992ApJ,Dall'Osso2011A&A,Rea2015ApJ,Stratta2018ApJ}.
The anti-correlation of $L_{\rm X}-T_{\rm a}$ and the three parameter
$L-T-E$ correlation are all argued to be clues supporting this
mechanism \citep{dainotti2008MNRAS,xu2012,Tang2019}. According to
this study, the upgraded $L-T-E$ correlation is $L_{\rm X} \propto
T_{\rm a}^{-0.99} E_{\gamma,\rm iso}^{0.86}$. Note that the power-law
index of $T_{\rm a}$ is close to $-1$ and the index of $E_{\gamma,\rm
iso}$ is close to 1. In other words, the correlation indicates
$L_{\rm X}  T_{\rm a} \propto E_{\gamma,\rm iso}$. Since $E_{\gamma,\rm
iso}$ itself is closely connected to the initial spin energy of
the millisecond magnetar, we see that the $L-T-E$ correlation does
support such an explanation for the plateau phase. Note that the
emergence of quasi-periodic oscillation signatures in the plateau
phase, which may be due to the precessing of a neutron star,
provides further support for this mechanism
\citep{Suvorov2021,zou2022MNRAS}.

Pseudo redshifts are reasonable estimate for the distances of GRBs
whose spectral redshifts were not measured. Especially, they are
useful for some statistical studies \citep{Zitouni2018Ap&SS}. Note
that the accuracy of the pseudo redshifts could be affected by
many factors, such as the error bars of the input parameters (i.e.
$F_{X}$, $S$, $T_{\rm a,obs}$, $\alpha_{\gamma}$, $\beta_{X}$),
the intrinsic scatter of the $L-T-E$ correlation, and even
possible deviation from the $L-T-E$ correlation by the target GRB
itself. Reducing the intrinsic dispersion ($\sigma _{\rm int }$)
of the empirical correlation could help improve the reliability of
the pseudo redshifts \citep{dainotti2011ApJ}.
\citet{Dainotti2011MNRAS} have built a gold sample to try to
derive a more compact relation, which is obviously a meaningful
exploration. More studies could be carried out in this aspect in
the future.


\section*{acknowledgements}
This study is supported by the National Natural Science Foundation
of China (Grant Nos. 12233002, 11873030, 12041306, 12147103,
U1938201), by National SKA Program of China No. 2020SKA0120300, by
the National Key R\&D Program of China (2021YFA0718500), and by
the science research grants from the China Manned Space Project
with NO. CMS-CSST-2021-B11. This work made use of data supplied by
the UK Swift Science Data Center at the University of Leicester.

\bibliography{sample631}{}
\bibliographystyle{aasjournal}

\begin{longrotatetable}
\begin{deluxetable*}{lcccccccc}
\tablenum{1} \tablecaption{Fitting Results and Relevant Parameters
of the Starting Sample. Note that only 36 GRBs are listed here,
which are new events as compared with the sample of
\citet{Tang2019}.  \label{tab:1}} \tablewidth{0pt} \tablehead{
\colhead{GRB Name} & \colhead{$z$} & \colhead{$T_{90}$} &
\colhead{$S$ (15-150 keV)} & \colhead{log $(T_{\rm a}/10^{3})$} &
\colhead{log $(L_{\rm X}/10^{47})$} & \colhead{log $(E_{\gamma,\rm
iso}/10^{53})$}&
\colhead{$\alpha_{1}$} & \colhead{$\alpha_{2}$}\\
\colhead{} & \colhead{} & \colhead{(s)} & \colhead{($\ 10^{-7} \rm erg~cm^{-2}$)} &
\colhead{(s)} & \colhead{($\rm erg~s^{-1}$)} & \colhead{(erg)}& \colhead{}& \colhead{}
}
\startdata
GRB 050318 & 1.44  & 32    & 10.8$\pm$0.77 & 0.63$\pm$0.18 & 0.08$\pm$0.29 & -1.27$\pm$0.03 & 0.56$\pm$0.21 & 2.21$\pm$0.18 \\
GRB 050820A & 2.612 & 26    & 34.4$\pm$2.42 & 1.55$\pm$0.17 & 0.1$\pm$0.45 & -0.68$\pm$0.03 & 0.95$\pm$0.02 & 1.51$\pm$0.05 \\
GRB 050822 & 1.434 & 103.4 & 24.6$\pm$1.72 & 0.74$\pm$0.2 & -0.22$\pm$0.12 & -0.74$\pm$0.03 & 0.14$\pm$0.09 & 1.15$\pm$0.05 \\
GRB 050915A & 2.5273 & 52    & 8.5$\pm$0.88 & 0.76$\pm$0.28 & -0.35$\pm$0.47 & -1.23$\pm$0.04 & 0.71$\pm$0.05 & 1.61$\pm$0.2 \\
GRB 050922B & 4.9   & 150.9 & 22.3$\pm$3.56 & 1.26$\pm$0.21 & 0.83$\pm$0.22 & 0.12$\pm$0.07 & -0.02$\pm$0.14 & 1.68$\pm$0.14 \\
GRB 051008 & 2.9   & 32    & 50.9$\pm$1.45 & 0.67$\pm$0.11 & 0.51$\pm$0.17 & -0.53$\pm$0.01 & 0.59$\pm$0.11 & 2.41$\pm$0.15 \\
GRB 060306 & 1.55  & 61.2  & 21.3$\pm$1.18 & -0.04$\pm$0.38 & 0.51$\pm$0.17 & -0.96$\pm$0.02 & 0.11$\pm$0.17 & 1.17$\pm$0.07 \\
GRB 060319 & 1.172 & 10.6  & 2.64$\pm$0.34 & 0.61$\pm$0.09 & -0.59$\pm$0.04 & -1.91$\pm$0.06 & -0.6$\pm$0.35 & 1.18$\pm$0.04 \\
GRB 060719 & 1.532 & 66.9  & 15$\pm$0.91 & 0.28$\pm$0.26 & 0.17$\pm$0.19 & -1.08$\pm$0.03 & 0.17$\pm$0.12 & 1.32$\pm$0.08 \\
GRB 070328 & 2.0627 & 75.3  & 90.6$\pm$1.79 & -0.77$\pm$0.04 & 2.49$\pm$0.02 & -0.39$\pm$0.01 & -0.27$\pm$0.1 & 1.53$\pm$0.02 \\
GRB 070419B & 1.9588 & 236.4 & 73.6$\pm$1.95 & 0.94$\pm$0.12 & 0.57$\pm$0.17 & -0.31$\pm$0.01 & 0.63$\pm$0.1 & 2.3$\pm$0.12 \\
GRB 070521 & 0.553 & 37.9  & 80.1$\pm$1.77 & 0.69$\pm$0.08 & -0.66$\pm$0.16 & -1.37$\pm$0.01 & 0.46$\pm$0.04 & 2.06$\pm$0.09 \\
GRB 071021 & 2.452 & 225   & 13$\pm$2 & 0.31$\pm$0.47 & 0$\pm$0.42 & -0.9$\pm$0.07 & 0.11$\pm$0.19 & 1.14$\pm$0.11 \\
GRB 080413B & 1.1   & 8     & 32$\pm$1 & 0.81$\pm$0.24 & -0.37$\pm$0.11 & -1.23$\pm$0.01 & 0.72$\pm$0.03 & 1.39$\pm$0.05 \\
GRB 080710 & 0.845 & 120   & 14$\pm$2 & 0.91$\pm$0.14 & -0.88$\pm$0.1 & -1.72$\pm$0.06 & 0.61$\pm$0.13 & 2.16$\pm$0.15 \\
GRB 081210 & 2.0631 & 146   & 18$\pm$2 & 1.95$\pm$0.06 & -1.38$\pm$0.11 & -1.01$\pm$0.05 & 0.64$\pm$0.05 & 3.71$\pm$0.54 \\
GRB 081222 & 2.7   & 24    & 48$\pm$1 & 0.66$\pm$0.17 & 0.63$\pm$0.27 & -0.62$\pm$0.01 & 0.88$\pm$0.02 & 1.65$\pm$0.07 \\
GRB 090404 & 3     & 84    & 30$\pm$1 & 0.7$\pm$0.07 & 0.83$\pm$0.06 & -0.03$\pm$0.01 & 0.12$\pm$0.05 & 1.44$\pm$0.05 \\
GRB 090426 & 2.609 & 1.2   & 1.8$\pm$0.3 & -1.28$\pm$0.23 & 1.29$\pm$0.09 & -1.59$\pm$0.07 & -0.38$\pm$0.5 & 1.09$\pm$0.06 \\
GRB 100724A & 1.288 & 1.4   & 1.6$\pm$0.2 & 0.25$\pm$0.32 & -0.62$\pm$0.27 & -2.18$\pm$0.05 & 0.69$\pm$0.09 & 1.5$\pm$0.17 \\
GRB 110801A & 1.858 & 385   & 47$\pm$3 & 0.59$\pm$0.22 & 0.17$\pm$0.34 & -0.46$\pm$0.03 & 0.7$\pm$0.09 & 1.8$\pm$0.14 \\
GRB 120907A & 0.97  & 16.9  & 6.7$\pm$1.1 & 1.13$\pm$0.14 & -1.14$\pm$0.14 & -1.85$\pm$0.07 & 0.72$\pm$0.02 & 1.39$\pm$0.08 \\
GRB 130505A & 2.27  & 88    & 210$\pm$10 & 0.49$\pm$0.13 & 1.6$\pm$0.11 & 0.31$\pm$0.02 & 0.39$\pm$0.13 & 1.63$\pm$0.04 \\
GRB 130609B & 1.3   & 210.6 & 160$\pm$0 & 0.79$\pm$0.05 & 0.33$\pm$0.08 & -0.39$\pm$0 & 0.85$\pm$0.03 & 2.84$\pm$0.14 \\
GRB 140419A & 3.956 & 94.7  & 160$\pm$0 & -0.18$\pm$0.28 & 2.07$\pm$0.16 & 0.15$\pm$0 & 0.3$\pm$0.19 & 1.48$\pm$0.07 \\
GRB 150314A & 1.758 & 14.79 & 220 $\pm$3 & 0.45$\pm$0.07 & 1.02$\pm$0.05 & -0.17$\pm$0.01 & -0.41$\pm$0.3 & 1.95$\pm$0.07 \\
GRB 161014A & 2.823 & 18.3  & 19$\pm$2 & 0.09$\pm$0.13 & 1.15$\pm$0.2 & -0.77$\pm$0.05 & 0.55$\pm$0.08 & 2.52$\pm$0.18 \\
GRB 190106A & 1.86  & 76.8  & 60$\pm$2 & 0.92$\pm$0.12 & 0.55$\pm$0.13 & -0.54$\pm$0.01 & 0.28$\pm$0.05 & 1.62$\pm$0.06 \\
GRB 190114A & 3.3765 & 66.6  & 8$\pm$1.2 & -0.3$\pm$0.22 & 1.32$\pm$0.09 & -0.67$\pm$0.07 & -0.09$\pm$0.19 & 1.25$\pm$0.1 \\
GRB 191011A & 1.722 & 7.37  & 3.3$\pm$0.4 & -0.73$\pm$0.18 & 0.57$\pm$0.09 & -1.63$\pm$0.05 & -0.24$\pm$0.24 & 1.36$\pm$0.09 \\
GRB 200829A & 1.25  & 13.04 & 510$\pm$10 & 0.47$\pm$0.06 & 1.16$\pm$0.08 & -0.09$\pm$0.01 & 0.87$\pm$0.01 & 1.89$\pm$0.04 \\
GRB 201221A & 5.7   & 44.5  & 18$\pm$2 & 0.54$\pm$0.1 & 0.3$\pm$0.17 & -0.5$\pm$0.05 & 0.585$\pm$0.05 & 2.92$\pm$0.75 \\
GRB 210210A & 0.715 & 6.6   & 9.8$\pm$0.8 & 0.46$\pm$0.1 & -0.18$\pm$0.1 & -1.99$\pm$0.04 & 0.29$\pm$0.07 & 1.8$\pm$0.11 \\
GRB 210610A & 3.54  & 13.62 & 10$\pm$1 & -0.8$\pm$0.16 & 1.5$\pm$0.06 & -0.97$\pm$0.04 & -0.39$\pm$0.18 & 1.12$\pm$0.07 \\
GRB 210722A & 1.145 & 50.2  & 25$\pm$2 & 0.59$\pm$0.28 & -0.12$\pm$0.32 & -1.19$\pm$0.03 & 0.63$\pm$0.09 & 1.73$\pm$0.17 \\
GRB 220521A & 5.6   & 13.55 & 8.1$\pm$0.8 & -0.94$\pm$0.12 & 2.1$\pm$0.06 & -0.39$\pm$0.04 & -0.56$\pm$0.21 & 1.68$\pm$0.14 \\
\enddata

\end{deluxetable*}
\end{longrotatetable}

\begin{longrotatetable}
\begin{deluxetable*}{lccccccccc}
\tablenum{2}
\tablecaption{Fitting Results and Relevant Parameters of 130 GRBs
in the Target Sample \label{tab:2}} \tablewidth{0pt} \tablehead{
\colhead{GRB Name} & \colhead{$z_{\rm est}$} & \colhead{$T_{90}$}
& \colhead{log $F_{X}$}& \colhead{$S$ (15-150 keV)} &\colhead{log
$(T_{\rm a}/10^{3})$} & \colhead{log $(L_{\rm X}/10^{47})$} &
\colhead{log $(E_{\gamma,\rm iso}/10^{53})$}&
\colhead{$\alpha_{1}$} & \colhead{$\alpha_{2}$}\\
\colhead{} & \colhead{} & \colhead{(s)} & \colhead{($\rm erg~cm^{-2}~s^{-1}$)}&
\colhead{($\ 10^{-7} \rm erg~cm^{-2}$)} &
\colhead{(s)} & \colhead{($\rm erg~s^{-1}$)} & \colhead{(erg)}& \colhead{}& \colhead{}
}
\startdata
 GRB 050607 & 14.21 & 26.4  & -11.89$\pm$0.24 & 5.92$\pm$0.55 & -0.4$\pm$0.46 & 1.96$\pm$0.24 & -0.03$\pm$0.04 & 0.21$\pm$0.19 & 1.3$\pm$0.14 \\
    GRB 050712 & 2.4   & 51.6  & -12.08$\pm$0.17 & 10.8$\pm$1.19 & 1.02$\pm$0.18 & -0.35$\pm$0.17 & -1.1$\pm$0.05 & 0.46$\pm$0.08 & 1.3$\pm$0.07 \\
    GRB 050713A & 2.54  & 124.7 & -11.08$\pm$0.22 & 51.1$\pm$2.12 & 0.53$\pm$0.17 & 0.76$\pm$0.22 & -0.37$\pm$0.02 & 0.53$\pm$0.04 & 1.43$\pm$0.05 \\
    GRB 050713B & 0.32  & 54.2  & -11.12$\pm$0.13 & 31.8$\pm$3.18 & 1.35$\pm$0.1 & -1.57$\pm$0.13 & -2.15$\pm$0.04 & 0.16$\pm$0.07 & 1.36$\pm$0.05 \\
    GRB 050726 & 1.05  & 49.9  & -10.95$\pm$0.35 & 19.4$\pm$2.12 & 0.41$\pm$0.22 & -0.17$\pm$0.35 & -1.59$\pm$0.05 & 0.66$\pm$0.1 & 2.01$\pm$0.17 \\
    GRB 060105 & 0.67  & 54.4  & -9.57$\pm$0.04 & 176$\pm$3.04 & 0.1$\pm$0.04 & 0.75$\pm$0.04 & -0.88$\pm$0.01 & 0.68$\pm$0.02 & 2$\pm$0.06 \\
    GRB 060109 & 0.98  & 115.4 & -11.18$\pm$0.08 & 6.55$\pm$1.03 & 0.47$\pm$0.1 & -0.4$\pm$0.08 & -1.8$\pm$0.07 & -0.21$\pm$0.11 & 1.68$\pm$0.1 \\
    GRB 060111B & 2.55  & 58.8  & -11.65$\pm$0.47 & 16$\pm$1.42 & 0.41$\pm$0.27 & 0.21$\pm$0.47 & -1.16$\pm$0.04 & 0.7$\pm$0.08 & 1.74$\pm$0.18 \\
    GRB 060204B & 1.44  & 139.4 & -11.25$\pm$0.32 & 29.5$\pm$1.78 & 0.58$\pm$0.25 & -0.05$\pm$0.32 & -1.26$\pm$0.03 & 0.6$\pm$0.08 & 1.71$\pm$0.11 \\
    GRB 060413 & 0.08  & 147.7 & -10.76$\pm$0.04 & 35.6$\pm$1.47 & 1.39$\pm$0.03 & -2.6$\pm$0.04 & -3.31$\pm$0.02 & -0.13$\pm$0.09 & 3.43$\pm$0.17 \\
    GRB 060428A & 0.04  & 39.5  & -11.79$\pm$0.08 & 13.9$\pm$0.78 & 2.17$\pm$0.07 & -4.18$\pm$0.08 & -4.26$\pm$0.02 & 0.48$\pm$0.02 & 1.68$\pm$0.06 \\
    GRB 060510A & 0.2   & 20.4  & -9.97$\pm$0.06 & 80.5$\pm$3.12 & 0.69$\pm$0.06 & -0.91$\pm$0.06 & -2.13$\pm$0.02 & 0.05$\pm$0.05 & 1.55$\pm$0.03 \\
    GRB 060712 & 3.65  & 26.3  & -11.89$\pm$0.27 & 12.4$\pm$2.17 & 0.2$\pm$0.42 & 0.79$\pm$0.27 & -0.72$\pm$0.08 & 0.12$\pm$0.16 & 1.24$\pm$0.12 \\
    GRB 060807 & 0.21  & 54    & -10.96$\pm$0.05 & 8.48$\pm$1.09 & 0.82$\pm$0.06 & -1.83$\pm$0.05 & -3.07$\pm$0.06 & -0.1$\pm$0.06 & 1.85$\pm$0.06 \\
    GRB 060813 & 0.98  & 16.1  & -10.45$\pm$0.11 & 54.6$\pm$1.4 & 0.4$\pm$0.15 & 0.22$\pm$0.11 & -1.15$\pm$0.01 & 0.61$\pm$0.04 & 1.62$\pm$0.06 \\
    GRB 061202 & 0.42  & 91.2  & -11.16$\pm$0.05 & 34.2$\pm$1.33 & 1.34$\pm$0.05 & -1.33$\pm$0.05 & -1.88$\pm$0.02 & 0.12$\pm$0.05 & 2.33$\pm$0.11 \\
    GRB 070107 & 4.78  & 347.3 & -12.62$\pm$0.07 & 51.7$\pm$2.64 & 1.65$\pm$0.04 & -0.18$\pm$0.07 & -0.18$\pm$0.02 & 0.9$\pm$0.01 & 2.84$\pm$0.3 \\
    GRB 070420 & 3.31  & 76.5  & -10.47$\pm$0.1 & 140$\pm$4.48 & 0.1$\pm$0.11 & 1.51$\pm$0.1 & 0.01$\pm$0.01 & 0.27$\pm$0.08 & 1.66$\pm$0.05 \\
    GRB 070628 & 0.82  & 39.1  & -10.66$\pm$0.1 & 35$\pm$2 & 0.7$\pm$0.11 & -0.14$\pm$0.1 & -1.22$\pm$0.02 & 0.22$\pm$0.06 & 1.51$\pm$0.08 \\
    GRB 071118 & 0.08  & 71    & -11.36$\pm$0.09 & 5$\pm$1 & 1.25$\pm$0.05 & -3.19$\pm$0.09 & -4.16$\pm$0.09 & 0.61$\pm$0.05 & 4$\pm$1.05 \\
    GRB 080212 & 0.92  & 123   & -11.11$\pm$0.23 & 29$\pm$3 & 0.62$\pm$0.38 & -0.43$\pm$0.23 & -1.66$\pm$0.04 & 0.3$\pm$0.28 & 1.63$\pm$0.16 \\
    GRB 080229A & 0.34  & 64    & -9.66$\pm$0.11 & 90$\pm$2 & 0.34$\pm$0.15 & -0.1$\pm$0.11 & -1.59$\pm$0.01 & 0.18$\pm$0.12 & 1.36$\pm$0.03 \\
    GRB 080328 & 6.3   & 90.6  & -9.57$\pm$0.07 & 94$\pm$2 & -1.16$\pm$0.27 & 3.05$\pm$0.07 & 0.36$\pm$0.01 & -0.16$\pm$0.4 & 1.19$\pm$0.07 \\
    GRB 080903 & 0.1   & 66    & -9.14$\pm$0.06 & 14$\pm$1 & -0.67$\pm$0.06 & -0.77$\pm$0.06 & -3.55$\pm$0.03 & -0.03$\pm$0.14 & 2.06$\pm$0.05 \\
    GRB 081230 & 16.32 & 60.7  & -12.06$\pm$0.36 & 8.2$\pm$0.8 & 0.16$\pm$0.25 & 1.65$\pm$0.36 & 0.24$\pm$0.04 & 0.47$\pm$0.06 & 1.46$\pm$0.15 \\
    GRB 090111 & 14.16 & 24.8  & -12.12$\pm$0.49 & 6.2$\pm$0.6 & 0.01$\pm$0.4 & 2.03$\pm$0.49 & 0.52$\pm$0.04 & 0.24$\pm$0.19 & 1.3$\pm$0.5 \\
    GRB 090518 & 1.35  & 6.9   & -11.21$\pm$0.26 & 4.7$\pm$0.4 & -0.01$\pm$0.54 & 0.05$\pm$0.26 & -1.82$\pm$0.04 & 0.19$\pm$0.17 & 1.18$\pm$0.15 \\
    GRB 090813 & 0.2   & 7.1   & -9.49$\pm$0.03 & 13$\pm$1 & -0.47$\pm$0.06 & -0.44$\pm$0.03 & -2.94$\pm$0.03 & -0.14$\pm$0.08 & 1.32$\pm$0.02 \\
    GRB 091130B & 2.33  & 112.5 & -12.05$\pm$0.12 & 13$\pm$1 & 1.24$\pm$0.17 & -0.22$\pm$0.12 & -0.69$\pm$0.03 & 0.13$\pm$0.09 & 1.4$\pm$0.11 \\
    GRB 100305A & 1.78  & 69.7  & -11.46$\pm$0.16 & 15$\pm$2 & 0.61$\pm$0.16 & -0.07$\pm$0.16 & -1.25$\pm$0.06 & 0.11$\pm$0.28 & 2.32$\pm$0.31 \\
    GRB 100508A & 0.18  & 52    & -11.51$\pm$0.05 & 7$\pm$1.1 & 1.37$\pm$0.04 & -2.58$\pm$0.05 & -3.32$\pm$0.07 & 0.33$\pm$0.04 & 3.29$\pm$0.3 \\
    GRB 100522A & 1.65  & 35.3  & -11.42$\pm$0.21 & 21$\pm$1 & 0.9$\pm$0.21 & -0.04$\pm$0.21 & -0.88$\pm$0.02 & 0.4$\pm$0.06 & 1.43$\pm$0.12 \\
    GRB 100619A & 7.08  & 97.5  & -11.67$\pm$0.3 & 45$\pm$1 & 0.55$\pm$0.2 & 1.38$\pm$0.3 & 0.37$\pm$0.01 & 0.69$\pm$0.03 & 1.34$\pm$0.09 \\
    GRB 100725B & 5.13  & 200   & -11.64$\pm$0.17 & 68$\pm$2 & 0.66$\pm$0.14 & 1.3$\pm$0.17 & 0.41$\pm$0.01 & 0.43$\pm$0.06 & 1.73$\pm$0.15 \\
    GRB 100727A & 2.89  & 84    & -11.38$\pm$0.06 & 12$\pm$1 & 0.38$\pm$0.18 & 0.58$\pm$0.06 & -0.76$\pm$0.04 & -0.2$\pm$0.11 & 1.09$\pm$0.11 \\
    GRB 101023A & 7.2   & 80.8  & -10.73$\pm$0.2 & 270$\pm$10 & -0.05$\pm$0.26 & 2.2$\pm$0.2 & 0.64$\pm$0.02 & 0.44$\pm$0.2 & 1.59$\pm$0.08 \\
    GRB 101024A & 1.25  & 18.7  & -10.09$\pm$0.06 & 15$\pm$1 & -0.51$\pm$0.09 & 0.81$\pm$0.06 & -1.51$\pm$0.03 & -0.37$\pm$0.18 & 1.37$\pm$0.05 \\
    GRB 101117B & 4.46  & 5.2   & -10.47$\pm$0.08 & 11$\pm$1 & -1.02$\pm$0.13 & 1.97$\pm$0.08 & -0.75$\pm$0.04 & -0.5$\pm$0.47 & 1.28$\pm$0.07 \\
    GRB 110102A & 4     & 264   & -10.53$\pm$0.08 & 165$\pm$3 & 0.22$\pm$0.09 & 1.75$\pm$0.08 & 0.44$\pm$0.01 & 0.24$\pm$0.05 & 1.48$\pm$0.04 \\
    GRB 110106B & 3.63  & 24.8  & -11.31$\pm$0.24 & 20$\pm$1 & 0.41$\pm$0.17 & 0.83$\pm$0.24 & -0.42$\pm$0.02 & 0.47$\pm$0.06 & 1.51$\pm$0.08 \\
    GRB 110210A & 5.51  & 233   & -11.84$\pm$0.08 & 9.6$\pm$1.9 & 0.48$\pm$0.37 & 0.68$\pm$0.08 & -0.52$\pm$0.09 & -0.23$\pm$0.22 & 0.93$\pm$0.14 \\
    GRB 110223A & 0.11  & 7     & -12.1$\pm$0.17 & 1.6$\pm$0.4 & 1.49$\pm$0.14 & -3.57$\pm$0.17 & -4.33$\pm$0.11 & 0.34$\pm$0.05 & 1.05$\pm$0.07 \\
    GRB 110312A & 1.41  & 28.7  & -11.61$\pm$0.13 & 8.2$\pm$1.3 & 0.97$\pm$0.1 & -0.43$\pm$0.13 & -1.25$\pm$0.07 & 0.32$\pm$0.05 & 1.17$\pm$0.07 \\
    GRB 110319A & 1.88  & 19.3  & -11.54$\pm$0.34 & 14$\pm$1 & 0.59$\pm$0.28 & -0.03$\pm$0.34 & -1.22$\pm$0.03 & 0.52$\pm$0.07 & 1.57$\pm$0.15 \\
    GRB 110411A & 3.88  & 80.3  & -10.97$\pm$0.21 & 33$\pm$2 & -0.51$\pm$0.52 & 1.82$\pm$0.21 & -0.34$\pm$0.03 & 0.04$\pm$0.25 & 1.26$\pm$0.18 \\
    GRB 110709A & 1.99  & 44.7  & -10.72$\pm$0.03 & 100$\pm$2 & 0.51$\pm$0.01 & 0.78$\pm$0.03 & -0.37$\pm$0.01 & 0.67$\pm$0.01 & 2.85$\pm$0.16 \\
    GRB 110709B & 1.65  & 55.6  & -11.76$\pm$0.15 & 94$\pm$2 & 1.65$\pm$0.1 & -0.44$\pm$0.15 & -0.49$\pm$0.01 & 0.87$\pm$0.03 & 2$\pm$0.1 \\
    GRB 110915A & 1.1   & 78.76 & -10.64$\pm$0.26 & 57$\pm$2 & 0.42$\pm$0.22 & 0.26$\pm$0.26 & -1.08$\pm$0.02 & 0.78$\pm$0.04 & 1.47$\pm$0.07 \\
    GRB 111129A & 0.05  & 7.6   & -10.77$\pm$0.2 & 1.8$\pm$0.4 & 0.33$\pm$0.25 & -2.93$\pm$0.2 & -4.91$\pm$0.1 & 0.27$\pm$0.1 & 1.32$\pm$0.07 \\
    GRB 120116A & 5.95  & 41    & -10.89$\pm$0.16 & 29$\pm$1 & -0.51$\pm$0.45 & 1.8$\pm$0.16 & -0.35$\pm$0.01 & -0.11$\pm$0.23 & 1.1$\pm$0.14 \\
    GRB 120308A & 1.33  & 60.6  & -11.26$\pm$0.11 & 12$\pm$1 & 0.8$\pm$0.07 & -0.35$\pm$0.11 & -1.36$\pm$0.04 & 0.56$\pm$0.04 & 2.61$\pm$0.2 \\
    GRB 120324A & 1.87  & 118   & -10.51$\pm$0.13 & 101$\pm$3 & 0.38$\pm$0.16 & 0.92$\pm$0.13 & -0.36$\pm$0.01 & 0.2$\pm$0.07 & 1.4$\pm$0.08 \\
    GRB 120701A & 3.99  & 13.8  & -11.81$\pm$0.35 & 14$\pm$1 & 0.32$\pm$0.14 & 0.42$\pm$0.35 & -1.02$\pm$0.03 & 0.85$\pm$0.04 & 1.76$\pm$0.18 \\
    GRB 120703A & 4     & 25.2  & -10.94$\pm$0.36 & 35$\pm$1 & 0.09$\pm$0.33 & 1.26$\pm$0.36 & -0.3$\pm$0.01 & 0.5$\pm$0.06 & 1.31$\pm$0.07 \\
    GRB 121031A & 0.96  & 226   & -11.06$\pm$0.38 & 78$\pm$2 & 1.16$\pm$0.22 & -0.28$\pm$0.38 & -0.87$\pm$0.01 & 0.39$\pm$0.08 & 1.55$\pm$0.34 \\
    GRB 121123A & 7.83  & 317   & -11.2$\pm$0.06 & 150$\pm$10 & 0.09$\pm$0.17 & 1.62$\pm$0.06 & 0.13$\pm$0.03 & -0.35$\pm$0.3 & 1.49$\pm$0.15 \\
    GRB 121217A & 1.23  & 778   & -10.89$\pm$0.08 & 62$\pm$3 & 0.91$\pm$0.11 & 0.05$\pm$0.08 & -0.77$\pm$0.02 & 0.12$\pm$0.08 & 1.52$\pm$0.07 \\
    GRB 130327A & 2.27  & 9     & -12.26$\pm$0.16 & 2.3$\pm$0.5 & 0.92$\pm$0.14 & -0.51$\pm$0.16 & -1.41$\pm$0.09 & 0.05$\pm$0.06 & 1.36$\pm$0.5 \\
    GRB 130504A & 0.2   & 50    & -10.6$\pm$0.22 & 10$\pm$1 & 0.47$\pm$0.23 & -1.55$\pm$0.22 & -3.14$\pm$0.04 & 0.27$\pm$0.16 & 1.78$\pm$0.22 \\
    GRB 130528A & 14.18 & 59.4  & -10.27$\pm$0.12 & 51$\pm$2 & -1.28$\pm$0.21 & 3.09$\pm$0.12 & 0.27$\pm$0.02 & -0.14$\pm$0.25 & 1.13$\pm$0.04 \\
    GRB 130803A & 3.33  & 44    & -11.05$\pm$0.43 & 15$\pm$1 & -0.18$\pm$0.6 & 1.23$\pm$0.43 & -0.65$\pm$0.03 & 0.44$\pm$0.11 & 1.03$\pm$0.1 \\
    GRB 131002A & 3.38  & 55.59 & -10.54$\pm$0.14 & 6.4$\pm$0.8 & -0.84$\pm$0.52 & 1.53$\pm$0.14 & -1.05$\pm$0.05 & -0.05$\pm$0.34 & 1.11$\pm$0.1 \\
    GRB 140108A & 1.12  & 94    & -10.57$\pm$0.18 & 70$\pm$2 & 0.7$\pm$0.19 & 0.26$\pm$0.18 & -0.76$\pm$0.01 & 0.36$\pm$0.1 & 1.64$\pm$0.15 \\
    GRB 140323A & 3.52  & 104.9 & -10.74$\pm$0.27 & 160$\pm$0 & 0.47$\pm$0.2 & 1.3$\pm$0.27 & 0.19$\pm$0 & 0.54$\pm$0.09 & 2$\pm$0.22 \\
    GRB 140709A & 1.57  & 98.6  & -11.12$\pm$0.23 & 53$\pm$2 & 0.89$\pm$0.22 & 0.05$\pm$0.23 & -0.79$\pm$0.02 & 0.44$\pm$0.08 & 1.59$\pm$0.15 \\
    GRB 140817A & 8.51  & 244   & -11.76$\pm$0.27 & 46$\pm$2 & 0.68$\pm$0.12 & 1.05$\pm$0.27 & 0.14$\pm$0.02 & 0.71$\pm$0.1 & 2.65$\pm$0.39 \\
    GRB 140916A & 0.14  & 80.1  & -11.28$\pm$0.05 & 17$\pm$3 & 1.53$\pm$0.06 & -2.53$\pm$0.05 & -3.07$\pm$0.08 & -0.2$\pm$0.08 & 2.3$\pm$0.18 \\
    GRB 141017A & 1.59  & 55.7  & -10.74$\pm$0.19 & 31$\pm$1 & 0.18$\pm$0.29 & 0.49$\pm$0.19 & -1.09$\pm$0.01 & 0.16$\pm$0.12 & 1.34$\pm$0.11 \\
    GRB 150201A & 0.004 & 26.1  & -9.13$\pm$0.03 & 7.8$\pm$1.2 & -0.43$\pm$0.07 & -3.64$\pm$0.03 & -6.63$\pm$0.07 & -0.15$\pm$0.12 & 1.3$\pm$0.02 \\
    GRB 150202A & 2.16  & 25.7  & -10.95$\pm$0.29 & 6.1$\pm$0.7 & -0.22$\pm$0.24 & 0.61$\pm$0.29 & -1.41$\pm$0.05 & 0.46$\pm$0.1 & 1.64$\pm$0.22 \\
    GRB 150203A & 6.79  & 25.8  & -11.66$\pm$0.28 & 9.1$\pm$0.6 & 0.22$\pm$0.18 & 1.14$\pm$0.28 & -0.28$\pm$0.03 & 0.19$\pm$0.15 & 2.47$\pm$0.9 \\
    GRB 150428B & 3.72  & 130.9 & -11.84$\pm$0.2 & 37$\pm$3 & 0.83$\pm$0.17 & 0.22$\pm$0.2 & -0.65$\pm$0.04 & 0.09$\pm$0.08 & 1.27$\pm$0.15 \\
    GRB 150430A & 5.04  & 107.1 & -10.47$\pm$0.13 & 69$\pm$5 & -0.41$\pm$0.15 & 2.05$\pm$0.13 & 0.05$\pm$0.03 & 0.17$\pm$0.18 & 1.97$\pm$0.14 \\
    GRB 160119A & 11.25 & 116   & -11.11$\pm$0.09 & 71$\pm$2 & -0.1$\pm$0.29 & 2.26$\pm$0.09 & 0.66$\pm$0.01 & -0.2$\pm$0.36 & 1.45$\pm$0.21 \\
    GRB 160504A & 1.79  & 53.9  & -11.49$\pm$0.13 & 7.3$\pm$0.9 & 0.65$\pm$0.16 & -0.14$\pm$0.13 & -1.28$\pm$0.05 & 0.08$\pm$0.07 & 1.52$\pm$0.18 \\
    GRB 160607A & 2.32  & 33.4  & -10.12$\pm$0.13 & 210$\pm$0 & 0.11$\pm$0.12 & 1.48$\pm$0.13 & -0.01$\pm$0 & 0.62$\pm$0.035 & 1.52$\pm$0.04 \\
    GRB 160630A & 0.73  & 29.5  & -10.31$\pm$0.09 & 12$\pm$1 & -0.08$\pm$0.16 & 0.03$\pm$0.09 & -1.93$\pm$0.04 & -0.03$\pm$0.12 & 1.25$\pm$0.07 \\
    GRB 160824A & 0.71  & 99.3  & -10.97$\pm$0.16 & 26$\pm$2 & 0.92$\pm$0.09 & -0.62$\pm$0.16 & -1.54$\pm$0.03 & 0.39$\pm$0.05 & 2.17$\pm$0.32 \\
    GRB 160905A & 0.33  & 64    & -9.57$\pm$0.06 & 150$\pm$2 & 0.35$\pm$0.06 & -0.03$\pm$0.06 & -1.5$\pm$0.01 & 0.64$\pm$0.02 & 1.58$\pm$0.03 \\
    GRB 161004B & 15.33 & 15.9  & -11.08$\pm$0.37 & 88$\pm$2 & -0.58$\pm$0.28 & 2.44$\pm$0.37 & 0.32$\pm$0.01 & 0.76$\pm$0.06 & 1.73$\pm$0.18 \\
    GRB 161202A & 3.09  & $\cdots$ & -11.26$\pm$0.59 & 86$\pm$3 & 0.43$\pm$0.38 & 1.16$\pm$0.59 & -0.02$\pm$0.02 & 0.7$\pm$0.08 & 1.72$\pm$0.24 \\
    GRB 161214B & 17.15 & 24.8  & -11.52$\pm$0.13 & 24$\pm$1 & -0.12$\pm$0.48 & 2.12$\pm$0.13 & 0.47$\pm$0.02 & -0.15$\pm$0.39 & 1.32$\pm$0.17 \\
    GRB 170317A & 1.84  & 11.94 & -11.16$\pm$0.23 & 13$\pm$1 & 0.07$\pm$0.24 & 0.25$\pm$0.23 & -1.5$\pm$0.03 & 0.39$\pm$0.09 & 1.66$\pm$0.2 \\
    GRB 170803A & 0.001 & 3.82  & -9.33$\pm$0.02 & 9.8$\pm$0.4 & -0.01$\pm$0.05 & -4.76$\pm$0.02 & -7.44$\pm$0.02 & 0.15$\pm$0.07 & 1.97$\pm$0.05 \\
    GRB 170906A & 1.86  & 88.1  & -9.1$\pm$0.03 & 320$\pm$0 & -0.74$\pm$0.08 & 2.4$\pm$0.03 & 0.08$\pm$0 & -0.58$\pm$0.29 & 1.76$\pm$0.05 \\
    GRB 171120A & 1.78  & 64    & -11.34$\pm$0.11 & 86$\pm$3 & 1.15$\pm$0.08 & -0.03$\pm$0.11 & -0.58$\pm$0.02 & 0.21$\pm$0.06 & 1.98$\pm$0.16 \\
    GRB 180411A & 3.16  & 77.5  & -10.36$\pm$0.14 & 120$\pm$0 & 0.1$\pm$0.15 & 1.57$\pm$0.14 & 0.08$\pm$0 & 0.08$\pm$0.1 & 1.45$\pm$0.09 \\
    GRB 180623A & 4.69  & 114.9 & -10.62$\pm$0.2 & 120$\pm$2 & 0.04$\pm$0.1 & 1.79$\pm$0.2 & 0.27$\pm$0.01 & 0.46$\pm$0.08 & 2.59$\pm$0.23 \\
    GRB 180626A & 6.44  & 30.07 & -11.07$\pm$0.15 & 48$\pm$2 & 0.11$\pm$0.17 & 1.64$\pm$0.15 & 0.18$\pm$0.02 & 0.24$\pm$0.06 & 1.44$\pm$0.1 \\
    GRB 190202A & 0.21  & 19.4  & -9.29$\pm$0.05 & 60$\pm$5 & -0.14$\pm$0.05 & -0.17$\pm$0.05 & -2.23$\pm$0.04 & -0.01$\pm$0.06 & 1.5$\pm$0.02 \\
    GRB 190203A & 1.95  & 96    & -10.31$\pm$0.12 & 150$\pm$5 & 0.4$\pm$0.07 & 0.96$\pm$0.12 & -0.29$\pm$0.01 & 0.9$\pm$0.02 & 2.11$\pm$0.1 \\
    GRB 190211A & 0.85  & 12.48 & -10.36$\pm$0.06 & 6.5$\pm$1.2 & -0.32$\pm$0.13 & 0.17$\pm$0.06 & -2.04$\pm$0.08 & -0.33$\pm$0.2 & 1.29$\pm$0.06 \\
    GRB 190519A & 11.08 & 45.58 & -10.2$\pm$0.17 & 180$\pm$10 & -0.62$\pm$0.21 & 2.78$\pm$0.17 & 0.66$\pm$0.02 & 0.23$\pm$0.09 & 1.41$\pm$0.1 \\
    GRB 190828B & 1.88  & 66.6  & -10.32$\pm$0.14 & 42$\pm$2 & -0.1$\pm$0.04 & 1.05$\pm$0.14 & -0.76$\pm$0.02 & -0.03$\pm$0.05 & 1.39$\pm$0.04 \\
    GRB 191004A & 1.94  & 2.44  & -10.49$\pm$0.14 & 13$\pm$1 & -0.31$\pm$0.14 & 0.86$\pm$0.14 & -1.22$\pm$0.03 & 0.54$\pm$0.06 & 1.8$\pm$0.12 \\
    GRB 200519A & 0.24  & 71.88 & -9.34$\pm$0.03 & 120$\pm$2 & 0.16$\pm$0.05 & -0.13$\pm$0.03 & -1.84$\pm$0.01 & 0.28$\pm$0.03 & 1.41$\pm$0.02 \\
    GRB 200711A & 1.63  & 29.39 & -10.66$\pm$0.24 & 53$\pm$3 & 0.43$\pm$0.19 & 0.59$\pm$0.24 & -0.69$\pm$0.02 & 0.47$\pm$0.07 & 1.852$\pm$0.19 \\
    GRB 200713A & 0.71  & 48.98 & -11.4$\pm$0.15 & 9$\pm$1.6 & 1.08$\pm$0.18 & -1.01$\pm$0.15 & -1.81$\pm$0.08 & 0.11$\pm$0.09 & 1.51$\pm$0.18 \\
    GRB 201209A & 4.99  & 48    & -10.2$\pm$0.06 & 140$\pm$10 & -0.23$\pm$0.15 & 2.21$\pm$0.06 & 0.45$\pm$0.03 & -0.24$\pm$0.1 & 1.08$\pm$0.08 \\
    GRB 201229A & 0.52  & 53.3  & -11.5$\pm$0.17 & 13$\pm$2 & 1.29$\pm$0.18 & -1.52$\pm$0.17 & -2.15$\pm$0.07 & 0.3$\pm$0.05 & 1.28$\pm$0.1 \\
    GRB 210104A & 0.54  & 32.06 & -9.39$\pm$0.15 & 88$\pm$2 & -0.14$\pm$0.28 & 0.66$\pm$0.15 & -1.27$\pm$0.01 & 0.17$\pm$0.17 & 1.41$\pm$0.12 \\
    GRB 210209A & 0.72  & 139.4 & -10.75$\pm$0.15 & 24$\pm$2 & 0.64$\pm$0.1 & -0.4$\pm$0.15 & -1.6$\pm$0.04 & 0.28$\pm$0.06 & 1.39$\pm$0.07 \\
    GRB 210306A & 0.18  & 9.12  & -9.5$\pm$0.05 & 57$\pm$1 & 0.09$\pm$0.06 & -0.54$\pm$0.05 & -2.41$\pm$0.01 & 0.31$\pm$0.03 & 1.64$\pm$0.05 \\
    GRB 210514A & 0.57  & 70.21 & -9.38$\pm$0.01 & 74$\pm$2 & -0.23$\pm$0.03 & 0.74$\pm$0.01 & -1.26$\pm$0.01 & -0.56$\pm$0.06 & 1.35$\pm$0.02 \\
    GRB 211129A & 0.58  & 113.01 & -10.72$\pm$0.11 & 23$\pm$2 & 0.63$\pm$0.11 & -0.52$\pm$0.11 & -1.75$\pm$0.04 & 0.09$\pm$0.08 & 1.52$\pm$0.08 \\
    GRB 220325A & 1.37  & 3.5   & -11.16$\pm$0.2 & 2.9$\pm$0.4 & -0.15$\pm$0.45 & -0.25$\pm$0.2 & -2.34$\pm$0.06 & 0.04$\pm$0.19 & 1.01$\pm$0.11 \\
    GRB 220408A & 0.84  & 17.25 & -10.47$\pm$0.21 & 12$\pm$1 & -0.03$\pm$0.26 & 0.04$\pm$0.21 & -1.86$\pm$0.04 & 0.18$\pm$0.15 & 1.49$\pm$0.11 \\
    GRB 220518A & 5.86  & 12.29 & -10.54$\pm$0.08 & 5.7$\pm$0.8 & -0.96$\pm$0.21 & 1.92$\pm$0.08 & -0.73$\pm$0.06 & -0.36$\pm$0.25 & 1.36$\pm$0.09 \\
    GRB 070220 & $\cdots$ & 129   & -11.55$\pm$0.17 & 104$\pm$2.33 & 1.3$\pm$0.09 & $\cdots$ & $\cdots$ & 0.97$\pm$0.04 & 2.67$\pm$0.33 \\
    GRB 090728 & $\cdots$ & 59    & -10.9$\pm$0.14 & 10$\pm$2 & 0.37$\pm$0.14 & $\cdots$ & $\cdots$ & 0.09$\pm$0.17 & 1.98$\pm$0.15 \\
    GRB 090807 & $\cdots$ & 140.8 & -11.92$\pm$0.18 & 22$\pm$2 & 1.23$\pm$0.12 & $\cdots$ & $\cdots$ & 0.14$\pm$0.12 & 2.52$\pm$0.27 \\
    GRB 090904A & $\cdots$ & 122   & -12.15$\pm$0.19 & 30$\pm$2 & 1.53$\pm$0.13 & $\cdots$ & $\cdots$ & 0.31$\pm$0.07 & 1.8$\pm$0.22 \\
    GRB 100614A & $\cdots$ & 225   & -12.57$\pm$0.1 & 27$\pm$2 & 2.32$\pm$0.07 & $\cdots$ & $\cdots$ & 0.31$\pm$0.07 & 2.9$\pm$0.5 \\
    GRB 110208A & $\cdots$ & 37.4  & -11.47$\pm$0.16 & 2.7$\pm$0.6 & 0.34$\pm$0.48 & $\cdots$ & $\cdots$ & 0.27$\pm$0.24 & 1.3$\pm$0.14 \\
    GRB 110315A & $\cdots$ & 77    & -11.52$\pm$0.17 & 41$\pm$2 & 1.15$\pm$0.12 & $\cdots$ & $\cdots$ & 0.74$\pm$0.05 & 1.6$\pm$0.09 \\
    GRB 110420A & $\cdots$ & 11.8  & -10.35$\pm$0.06 & 59$\pm$2 & 0.35$\pm$0.11 & $\cdots$ & $\cdots$ & 0$\pm$0.08 & 1.2$\pm$0.04 \\
    GRB 121001A & $\cdots$ & 147   & -10.17$\pm$0.65 & 17$\pm$2 & 0.03$\pm$0.65 & $\cdots$ & $\cdots$ & 0.8$\pm$0.16 & 1.7$\pm$0.19 \\
    GRB 130315A & $\cdots$ & 233.4 & -12.21$\pm$0.2 & 49$\pm$2 & 1.49$\pm$0.15 & $\cdots$ & $\cdots$ & 0.04$\pm$0.17 & 1.88$\pm$0.39 \\
    GRB 130527A & $\cdots$ & 44    & -9.92$\pm$0.1 & 120$\pm$4 & -0.09$\pm$0.17 & $\cdots$ & $\cdots$ & -0.08$\pm$0.29 & 1.81$\pm$0.15 \\
    GRB 130725B & $\cdots$ & 10    & -10.33$\pm$0.07 & 3.9$\pm$0.4 & -0.26$\pm$0.17 & $\cdots$ & $\cdots$ & -0.22$\pm$0.15 & 1.07$\pm$0.05 \\
    GRB 140730A & $\cdots$ & 41.3  & -11.78$\pm$0.41 & 2.8$\pm$0.5 & 0.93$\pm$0.36 & $\cdots$ & $\cdots$ & 0.35$\pm$0.1 & 1.45$\pm$0.32 \\
    GRB 150626A & $\cdots$ & 144   & -11.56$\pm$0.25 & 18$\pm$2 & 1.1$\pm$0.37 & $\cdots$ & $\cdots$ & -0.01$\pm$0.13 & 1.18$\pm$0.3 \\
    GRB 150817A & $\cdots$ & 38.8  & -10.11$\pm$0.31 & 59$\pm$1 & 0.13$\pm$0.39 & $\cdots$ & $\cdots$ & 0.08$\pm$0.18 & 1.3$\pm$0.08 \\
    GRB 171102B & $\cdots$ & 17.8  & -11.17$\pm$0.31 & 9.3$\pm$1.4 & 0.95$\pm$0.21 & $\cdots$ & $\cdots$ & 0.35$\pm$0.1 & 1.97$\pm$0.38 \\
    GRB 180620A & $\cdots$ & 23.16 & -10.68$\pm$0.06 & 58$\pm$2 & 0.94$\pm$0.03 & $\cdots$ & $\cdots$ & 0.06$\pm$0.06 & 3.32$\pm$0.24 \\
    GRB 180925A & $\cdots$ & 81.7  & -10.76$\pm$0.33 & 41$\pm$3 & -0.05$\pm$0.71 & $\cdots$ & $\cdots$ & 0.35$\pm$0.12 & 1.55$\pm$0.25 \\
    GRB 200906A & $\cdots$ & 70.9  & -11.51$\pm$0.39 & 28$\pm$1 & 0.84$\pm$0.46 & $\cdots$ & $\cdots$ & 0.26$\pm$0.12 & 1.14$\pm$0.2 \\
    GRB 220319A & $\cdots$ & 6.44  & -11.3$\pm$0.13 & 2.3$\pm$0.4 & 0.09$\pm$0.08 & $\cdots$ & $\cdots$ & 0.03$\pm$0.17 & 3.44$\pm$0.53 \\
    GRB 220403B & $\cdots$ & 27    & -11.3$\pm$0.17 & 37$\pm$1 & 0.94$\pm$0.19 & $\cdots$ & $\cdots$ & 0.06$\pm$0.07 & 1.32$\pm$0.13 \\
    GRB 220430A & $\cdots$ & 43.1  & -10.61$\pm$0.23 & 440$\pm$5 & 0.84$\pm$0.13 & $\cdots$ & $\cdots$ & 0.88$\pm$0.03 & 1.72$\pm$0.07 \\
\enddata

\end{deluxetable*}
\end{longrotatetable}



\end{CJK*}
\end{document}